\documentclass[sigconf,nonacm,10pt]{acmart}
\AtBeginDocument{%
	}

\usepackage{comment}
\usepackage{tikz}
\usepackage{xcolor}
\usepackage{xspace}
\usepackage{multirow}
\usepackage{subfigure}
\usepackage{algorithm,algpseudocode}
\usepackage{fancyhdr}


\renewcommand\footnotetextcopyrightpermission[1]{} 
\newcommand{\name}{Arax\xspace}
\newcommand{\auto}{Autotalk\xspace}

\newcommand*\circled[1]{\tikz[baseline=(char.base)]{\node[shape=circle,fill=gray,inner sep=0.4pt] (char) {\textcolor{white}{#1}};}}

\title{Arax: A Runtime Framework for Decoupling Applications from Heterogeneous Accelerators}
\author{Manos Pavlidakis$^{1,2}$, Stelios Mavridis$^{1}$, Antony Chazapis$^1$, Giorgos Vasiliadis$^1$, and Angelos Bilas$^{1,2}$}
\affiliation{
\institution{{$^1$Institute of Computer Science (ICS), Foundation for Research and Technology - Hellas (FORTH)}\country{Greece}}
\institution{{$^2$Computer Science Department, University of Crete}\country{Greece}}
}
\email{{manospavl, mavridis, chazapis, gvasil, bilas}@ics.forth.gr}

\newif\ifanswers

\setcopyright{none}
\renewcommand\footnotetextcopyrightpermission[1]{} 

\begin{document}

\rhead{\footnotesize Pavlidakis et al.}
\begin{abstract}

Today, using multiple heterogeneous accelerators efficiently from applications and high-level frameworks, such as TensorFlow and Caffe, poses significant challenges in three respects: (a) sharing accelerators, (b) allocating available resources elastically during application execution, and (c) reducing the required programming effort.

In this paper, we present \name{}, a runtime system that decouples applications from heterogeneous accelerators within a server. First, \name maps application tasks dynamically to available resources, managing all required task state, memory allocations, and task dependencies. As a result, \name{} can share accelerators across applications in a server and adjust the resources used by each application as load fluctuates over time. Additionally, \name{} offers a simple API and includes \auto, a stub generator that automatically generates stub libraries for applications already written for specific accelerator types, such as NVIDIA GPUs. Consequently, \name applications are written once without considering physical details, including the number and type of accelerators. 

Our results show that applications, such as Caffe, TensorFlow, and Rodinia, can run using \name{} with minimum effort and low overhead compared to native execution, about 12\% (geometric mean). \name{} supports efficient accelerator sharing, by offering up to 20\% improved execution times compared to NVIDIA MPS, which supports NVIDIA GPUs only. \name{} can transparently provide elasticity, decreasing total application turn-around time by up to 2$\times$ compared to native execution without elasticity support.

\end{abstract}

\maketitle
\section{Introduction}
\label{sec:introduction}

The increasing need for high performance at low energy consumption has resulted in the proliferation of heterogeneous accelerators, such as GPUs, FPGAs, and TPUs~\cite{fpgas_microsoft,tpus,starpu,het-accels,het-accels3,het-accels4}.
Recent estimates~\cite{het-accels4, starpu, batra2018artificial,extreme_het,ava} indicate that by 2030 servers will include a plethora of processing units and specialized accelerators~\cite{origami, diannao, pudiannao, simba}. This trend poses significant challenges in how applications and higher-level frameworks, such as TensorFlow~\cite{tensorflow} and Caffe~\cite{caffe}, can fully utilize the capacity of heterogeneous accelerators.

Today, a large percentage of applications or frameworks is \emph{statically bound to specific accelerators throughout their execution.} Many applications are directly written for one accelerator type, e.g.,  NVIDIA GPUs, to allow for device-specific optimizations. Over the last years, unified programming models, e.g., SYCL~\cite{sycl} and oneAPI~\cite{oneapi}, aim to offer portability to different accelerator types. However, applications are still required to explicitly select the desired accelerators during initialization and prior to starting their execution. As a result, each application execution is still bound to a specific set of accelerators or accelerator types that cannot change at runtime. This results in poor resource and application efficiency in two ways: (a) reduced sharing of resources and (b) lack of adaptation over time.

First, existing resource assignment techniques fully allocate accelerators to a single application. Although practical, this exclusive assignment creates significant \emph{load imbalance} in heterogeneous setups with multiple accelerators and results in resource under-utilization. Existing ti\-me-sharing appro\-aches~\cite{gandiva, antman, ava, dcuda} cannot address this issue effectively, e.g., in cases where an application cannot fully utilize an accelerator during its time slice. Spatial sharing, on the other hand, has the potential to increase resource utilization. However, existing approaches, such as NVIDIA MPS~\cite{mps}, are limited to specific accelerator types and require applications to perform manual task assignment and data placement.

Second, resources assigned to each application remain fixed throughout its execution. However, applications often exhibit dynamic behavior and fluctuating load requirements~\cite{dcuda,gandiva}. Given that it is difficult to estimate the resource demands of applications accurately and statically assign resources to each application, the \emph{lack of elasticity mechanisms} results in application under- or over-provisioning and eventually to poor resource utilization.

\begin{table}[t]
\centering
\scalebox{0.8}{
\begin{tabular}{|c|c|c|c|c|c|c|}
\hline
\textbf{Capabilities}                                                                           &
\multicolumn{1}{l|}{\textbf{\begin{tabular}[c]{@{}c@{}}MPS\\ \cite{mps}\end{tabular}}}          &   
\multicolumn{1}{l|}{\textbf{\begin{tabular}[c]{@{}c@{}}StarPU\\ \cite{starpu}\end{tabular}   }} &   
\multicolumn{1}{l|}{\textbf{ \begin{tabular}[c]{@{}c@{}}Gandiva\\ \cite{gandiva}\end{tabular}}} &   
\multicolumn{1}{l|}{\textbf{\begin{tabular}[c]{@{}c@{}}DCUDA\\ \cite{dcuda}\end{tabular}}}      &
\textbf{\begin{tabular}[c]{@{}c@{}}AvA\\ \cite{ava}\end{tabular}}                               &   
\multicolumn{1}{l|}{\textbf{Arax}} \\ \hline
\begin{tabular}[c]{@{}c@{}}Heterogeneity\end{tabular}       &- &\checkmark &-  &- &\checkmark &\checkmark\\\hline
\begin{tabular}[c]{@{}c@{}}Spatial sharing\end{tabular}          &\checkmark &-  &- &- &- &\checkmark\\\hline
\begin{tabular}[c]{@{}c@{}}Dynamic\\resource assign.\end{tabular}&- &- &\checkmark &\checkmark &- &\checkmark\\\hline
\begin{tabular}[c]{@{}c@{}}Reducing effort\end{tabular}    &- &- &- &- &\checkmark &\checkmark\\\hline
\end{tabular}
}
\caption{Capabilities of \name{} vs. state-of-the-art approaches.}
\label{tab:compare_with_prior}
\end{table}

In this paper, we present \name, a runtime system that decouples applications from heterogeneous accelerators \emph{within} a single server. Our approach is based on RPC, a mechanism that is proven to be very successful in decoupling complex software stacks, using clear and conceptually simple boundaries. The client-side stubs of \name allow applications to be written once using a simple API, without considering any low-level details, such as the number or type of accelerators. The core component of \name is a backend service, the \name server, that dynamically maps application tasks and data to available accelerators at runtime. This enables spatial accelerator sharing and adjusts resources at runtime. Last but not least, \name includes a stub generator (\auto) that reduces porting effort for existing accelerator-enabled applications. Table~\ref{tab:compare_with_prior} summarizes the main capabilities of \name, compared to state-of-the-art approaches. The whole \name ecosystem is available at GitHub\footnote{https://github.com/CARV-ICS-FORTH/arax}.

The RPC-based approach of \name allows \textbf{decoupling accelerators from applications}. \name applications do not need to perform accelerator selection, memory allocation, or task assignment operations; all are handled transparently by \name. This approach allows \name{} to perform memory allocations lazily and only when the actual task assignment occurs. To improve accelerator utilization  while ensuring application performance \name{} provides three capabilities:

(a) \textbf{Spatial sharing} that manages existing mechanisms in heterogeneous accelerators, transparently, and across all applications in a server. We use asynchronous host-threads to issue tasks to GPU streams and FPGA command queues. Regarding FPGAs, \name loads bitstreams with multiple kernels that need to be collocated in the same FPGA. The advantage of our approach is that it moves all the related management from individual applications to the shared \name{} runtime and can make decisions across all applications.

(b) \textbf{Elasticity and dynamic resource assignment} to applications at runtime. To achieve this, \name requires fine-grain access to application tasks and their data. \name uses asynchronous operations to issue independent tasks across different accelerators, while ensuring that tasks with dependencies execute in-order.

(c) \textbf{Live-migration} that moves application tasks across heterogeneous accelerators. Unlike existing approaches, our migration mechanism does not require application modifications or specialized accelerator support. \name uses task arguments to keep track of the data used by each task and transfers only relevant data upon task migration. Although arbitrary pointers may result in moving large amounts of memory, our approach is adequate to support real applications, such as TensorFlow and Caffe.

Finally, \name includes \textbf{\auto}, a generator that creates stubs for a given accelerator API based on a description of the target API. Applications are then linked dynamically with the stub library that internally calls the \name API. Currently, \auto generates stubs for a subset of CUDA that can support Caffe and TensorFlow.

We evaluate \name using Caffe, TensorFlow, and Rodinia. Our results show that \name applications can run without any modifications at low overhead---up to 12\% compared to native---when other approaches, i.e., AvA~\cite{ava}, result in up to 30\% overhead for the same applications. In addition, \name provides elasticity, decreasing total application turnaround time by 2$\times$ compared to native execution without elasticity support. Our migration mechanism adds 7\% overhead compared to standalone execution. Finally, our sharing mechanism provides up to 20\% improvement in total execution time compared to NVIDIA MPS.

The main contributions of this paper are:

\begin{enumerate}
\item We propose an RPC-based approach to decouple applications from heterogeneous accelerators within servers.
\item We present a mechanism for spatial sharing of heterogeneous accelerators and dynamic and transparent assignment of tasks to accelerators.
\item We present an application live-migration mechanism that reduces data movement based on data ownership by tasks.
\item We present a stub generator that allows existing applications to use \name{} with minimal effort and demonstrate our approach with Caffe and TensorFlow.
\item We demonstrate and evaluate \name in an accelerator-rich server environment, using GPUs, FPGAs, and CPUs, with Caffe, TensorFlow, and Rodinia.
\end{enumerate}

\section{Design}
\label{sec: design}
\begin{figure}[t]
	\includegraphics[width=1.02\columnwidth]{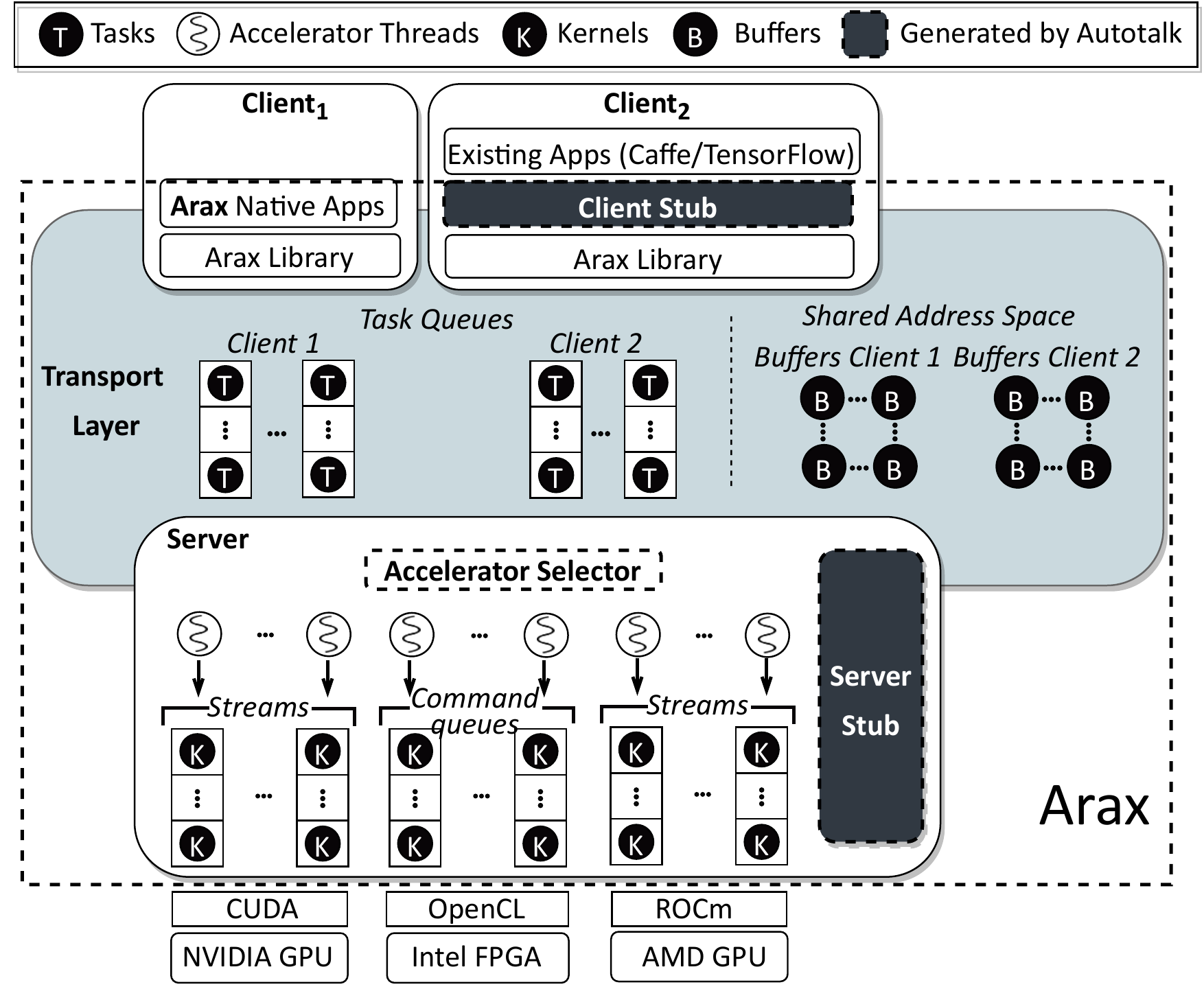}
	\caption{\name{} high-level overview. The main components of \name are: Clients, Server, Transport layer, and \auto.}
	\label{fig:rush_stack}
\end{figure}

Figure~\ref{fig:rush_stack} shows a high-level overview of \name. Applications use the \name API to access available accelerators, regardless of their types. Applications create task queues and issue tasks, providing their data in the form of \name buffers. Tasks and buffers are being transported to the \name server via a transport layer over shared memory, mapped to both the application and server address spaces. The \name server assigns dynamically and asynchronously application tasks to accelerators, managing accelerator streams and command queues, maintaining task ordering, and handling data dependencies. Finally, \name{}'s stub generator, \auto, allows generating the stub library automatically for a particular accelerator API, given a description file of the API calls. Next, we discuss each component of \name in more detail.

\subsection{Client}
\label{subsec:client}

\name{} provides three basic abstractions: (a) \emph{tasks}, (b) \emph{task buffers}, and (c) \emph{task queues}. Table~\ref{tab:api_calls} shows an overview of the main \name API calls.

\paragraph{\textbf{Tasks}} 
A task can be either a compute or a transfer task. A compute task is an accelerator kernel, while a transfer task is a data transfer between the host and the accelerator. Both tasks are executed without interruption and are asynchronous. \name{} provides synchronization primitives to allow applications to wait for their completion. A compute task takes the kernel name and its corresponding arguments as parameters, i.e., inputs, outputs, and arguments required from a kernel. The kernel name is associated with the actual kernel at the server (\S\ref{subsec:server}). Unlike existing accelerator APIs, task arguments do not include accelerator-specific information, such as thread number or thread block size. The parameters for a transfer task include the task buffers provided by \name and any data from the application address space. 

\paragraph{\textbf{Task buffers}} 
A buffer represents the input and output data of a task. Multiple tasks or applications can operate on the same buffer concurrently. It is important to note that \name decouples the accelerator memory management from applications using a lazy memory allocation strategy. When an application requests memory, \name stores the requested allocation size but does not allocate this memory on the accelerator (\S\ref{subsec:server}). The actual allocation will be performed only after the task is successfully assigned to an accelerator. In the meantime, applications can continue issuing tasks since buffers are implemented as opaque types in the shared memory. For all allocations in the shared memory, we use the Doug Lea allocator. This abstraction hides accelerator memory, and applications are unaware of which accelerator hosts their data.  
 
\paragraph{\textbf{Task queues}} 
Applications issue tasks to task queues, similar to existing programming models, e.g., CUDA/ROCm streams and OpenCL command queues. The main difference of \name is that these queues are not assigned directly to an accelerator. Instead, \name is responsible for assigning them to one or more accelerators at runtime (\S\ref{subsec:server}), while ensuring that asynchronous tasks will be executed in-order. Each task queue holds tasks with dependencies. To denote independent sets of work, applications need to acquire different task queues. This approach works well for the ML frameworks we examine due to the inherent serialization of NN layers.

\begin{table}[t]
\centering
\scalebox{0.8}{
\begin{tabular}{|c|c|c|}
\hline
\textbf{Abstraction}          & \textbf{API call}   & \textbf{Description} \\ \hline
\multirow{2}{*}{Tasks}        & a\_issue()          & Issue a task         \\ \cline{2-3}
                              & a\_wait()           & Wait for a task      \\ \hline
\multirow{3}{*}{Task Buffers} & a\_allocate()       & Allocate Buffer      \\ \cline{2-3}
                              & a\_free()           & Free Buffer          \\ \cline{2-3}
                              & \begin{tabular}[c]{@{}c@{}}a\_sync\_to(), a\_sync\_from()\end{tabular} & Transfer Data\\ \hline
\multirow{2}{*}{Task Queues}  & a\_acquire()        & Acquire a queue      \\ \cline{2-3}
                              & a\_release()        & Release a queue      \\ \hline
\end{tabular}
}
\caption{Methods of \name{} API.}
\label{tab:api_calls}
\end{table}
\subsection{Server}
\label{subsec:server}

The \name server is responsible for maintaining task issue order and managing data dependencies while performing dynamic task assignment and data placement to accelerators. These mechanisms allow \name{} to provide efficient spatial sharing and elastic allocation of resources.

\paragraph{\textbf{Spatial Sharing}}
\label{subsec:sharing}
The spatial sharing mechanism of \name is based on streams/command queues and host-threads (\name accelerator threads). In particular, to execute kernels in parallel, the server spawns multiple threads per physical accelerator. Each accelerator thread internally creates different streams (CUDA and ROCm) or command queues (OpenCL). The design of spatial sharing in \name can support advanced task assignment policies that do not rely on low-level accelerator-specific APIs. To enable spatial sharing for NVIDIA GPUs, we require a single context; thus, the \name server is implemented as a single process for all accelerators. Regarding FPGAs, the \name server loads a bitstream that contains multiple kernels, similar to Vinetalk~\cite{mavridis2017vinetalk}. The server can select and load the appropriate bitstream to serve each task. 

\paragraph{\textbf{Application migration}} Even when accelerators are sha\-red, there can be load imbalances. \name offers an application migration mechanism to correct load imbalances. This migration mechanism can move application tasks and their data across heterogeneous accelerators. The migration mechanism cannot stop a task during execution. Instead, it waits for the task to finish and moves any pending tasks and their data to another accelerator. There are \textit{three challenges} that our migration mechanism needs to tackle:  
\begin{figure}[t]
\includegraphics[width=\columnwidth]{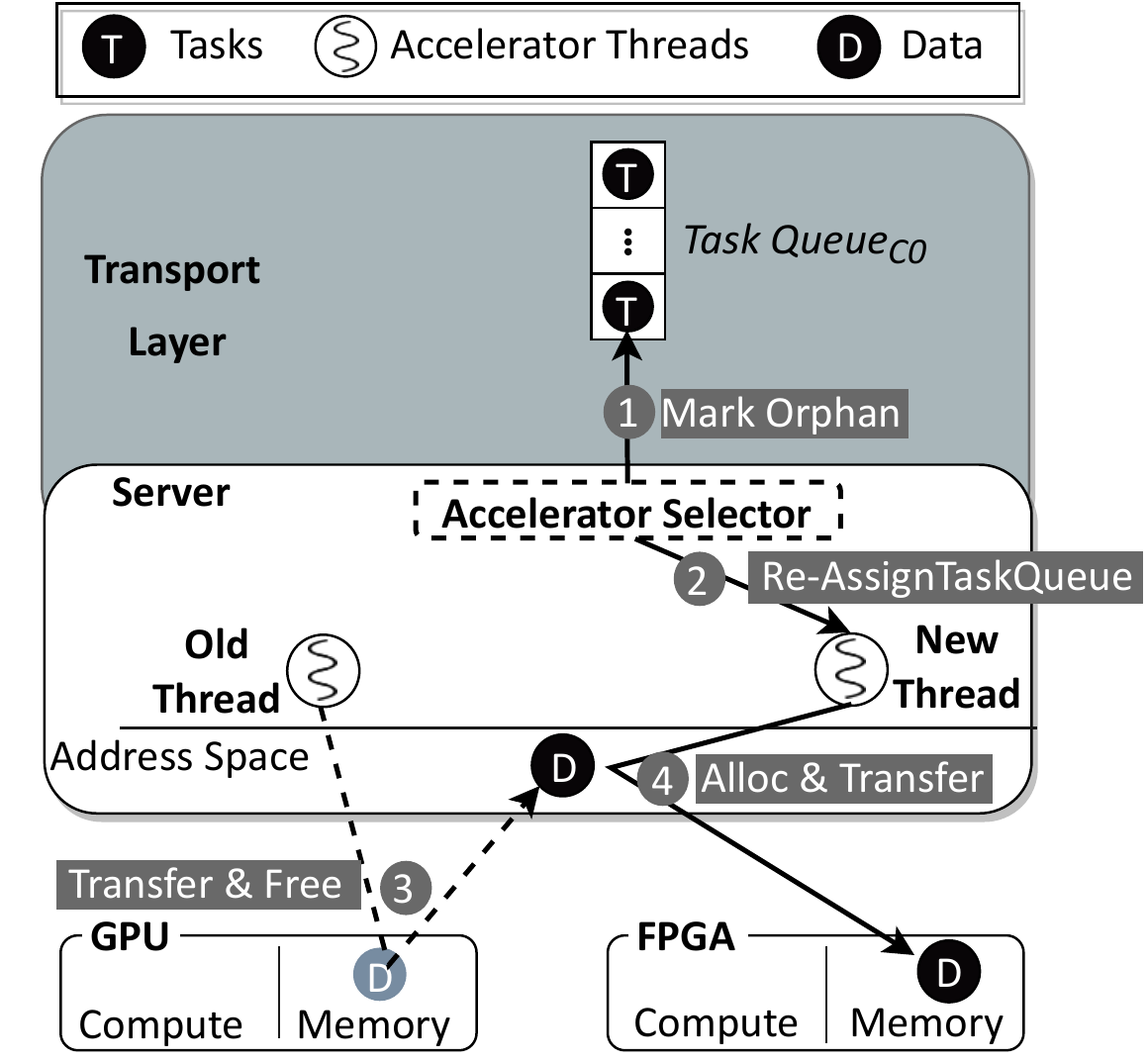}
\caption{The steps required for an application migration.  The task queue is marked orphan (1) and reassigned to a new thread (2). The relevant data are then transferred to the new accelerator via the server memory (3,4).
	}
\label{fig:migration}
\end{figure}

\textit{(i) Migrate an application without interrupting its execution.} \name offers task queues to applications to issue their tasks. The \name server stops and resumes the execution of a task queue, and thus it does not affect the execution of the application. In particular, \name performs the following steps: (a) The server marks this task queue as an orphan (Figure~\ref{fig:migration}; step \circled{1}). At this point, accelerator threads cannot launch tasks from this task queue. (b) Since then, there could have been tasks issued for execution; the server waits for them to finish before re-assigning this task queue to a different accelerator thread (Figure~\ref{fig:migration}; step \circled{2}). (c) From here on, any remaining task from this particular task queue will be invoked to the new accelerator. We note that, during the migration, the application continues issuing tasks to its task queues.

\textit{(ii) Move only the data of the migrated task.} The server should move only the data required from the migrated task and not all the application state. Existing checkpoint approaches~\cite{gandiva, gandivafair} migrate all the application state, which involves transfers in the range of gigabytes. The \name server maintains metadata for each task and is aware of the data required. After assigning the task queue to a new accelerator thread, the server instructs the previous accelerator thread to copy the task data from its accelerator memory to the server memory and free the corresponding allocations (Figure~\ref{fig:migration}; step \circled{3}). The server then notifies the new accelerator thread to allocate and copy that data from the server's memory (Figure~\ref{fig:migration}; step \circled{4}) using the native accelerator API. We note that the server memory is an intermediate buffer to transfer data across different accelerators. As part of our future work, we plan to eliminate this extra copy using accelerator-to-accelerator transfers, at least for the cases supported~\cite{gpudirect}.

\textit{(iii) Migrate the most recent version of the data.} Before a data migration, we must ensure that the data required from the migrated task(s) are up-to-date. To achieve that, the server allows only one valid copy of the data (at any given time) to the distinct accelerator memories in multi-accelerator setups. 

\paragraph{\textbf{Dynamic task assignment}}
\label{subsec:dynamic}

The server assigns the incoming task queues to the underlying accelerators. Individual tasks from the same task queue can be assigned to different accelerators. This assignment involves task and data migrations for tasks with dependencies. When the server detects an unassigned, non-empty task queue, it assigns it to an accelerator using a round-robin policy (default). Advanced assignment policies can be implemented with relatively low effort. This is facilitated by the fact that \name already collects information regarding the memory footprint of each task, the number of tasks per accelerator, and the data ownership.

As a proof of concept that our accelerator selector can host advanced assignment policies, we also implement an elastic assignment policy. This policy is essential to handle load fluctuations or data bursts by performing dynamic task assignments. The server keeps track of the assigned task queues per accelerator and knows the owner of each task queue. Consequently, the accelerator selector can increase/decrease the accelerators assigned to an application based on the load. 

For instance, lets assume that we have a low-priority application with two task queues, i.e., \textit{task queue1} and \textit{task queue2}. Initially, both task queues are assigned to the same accelerator. When the accelerator selector detects idle accelerators, it expands the resources used by the low-priority application by assigning \textit{task queue2} to the idle accelerator. Reversely, when another high-priority application arrives, the server shrinks the accelerators used from the low-priority application by moving \textit{task queue2} to the accelerator where \textit{task queue1} executes. This re-assignment requires moving the application state between accelerators, i.e., application migration. Consequently, the high-priority application can make exclusive use of the idle accelerator.

\begin{figure}[t]
\includegraphics[width=0.9\columnwidth]{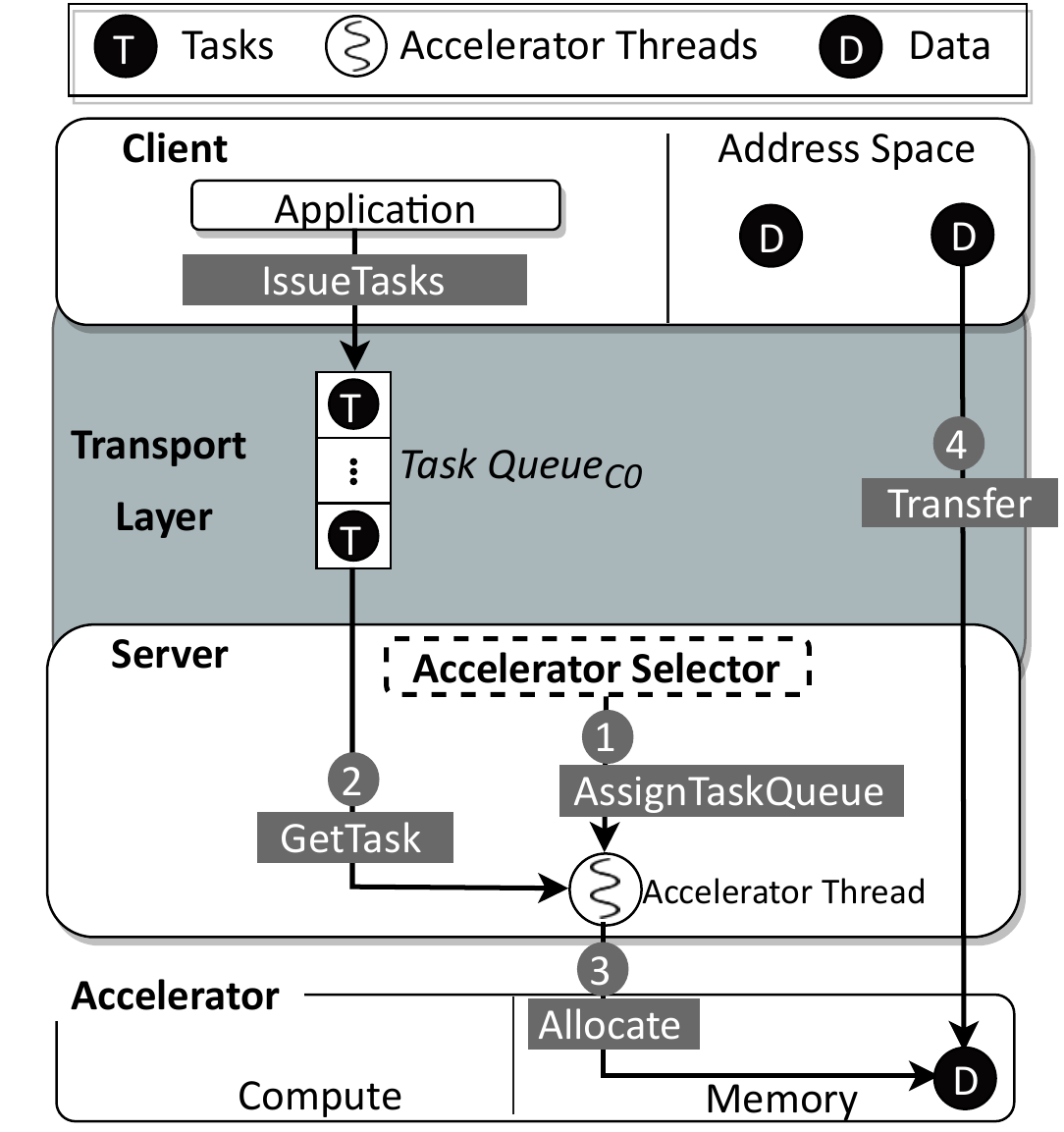}
\caption{\name dynamic task assignment. Application issues tasks to a task queue.  Initially, the task queue is assigned to
  an accelerator (1), then the accelerator thread gets a task (2). It allocates accelerator memory for that data (3) and
  copies the data from the application (4).}
\label{fig:rush_controller}
\end{figure}

To perform memory management, the server maintains internally a mapping of the allocated buffers per task queue and their corresponding sizes. We note that the actual memory allocation is performed only after its corresponding task queue has been assigned to a physical accelerator (Figure~\ref{fig:rush_controller}; step \circled{1}). After the selection of the physical accelerator, the thread of that accelerator gets a task from the task queue (Figure~\ref{fig:rush_controller}; step \circled{2}) and checks if any memory has already been allocated in that particular accelerator memory. If not, it performs the actual allocation (Figure~\ref{fig:rush_controller}; step \circled{3}) and keeps a reference to that memory segment so that it can be used for deallocation purposes. After that, the accelerator thread can issue the task to the accelerator. If the task is a data transfer, the accelerator thread copies the data from the client address space to the accelerator memory (Figure~\ref{fig:rush_controller}; step \circled{4}).

To support different accelerator types, the server spawns separate accelerator threads. Each thread uses the accelerator's native API to communicate with that particular accelerator. Currently, \name supports NVIDIA GPUs using CUDA, Intel Altera FPGAs using OpenCL, and AMD GPUs using ROCm. When receiving a compute task, the accelerator thread uses the kernel name---passed as a task parameter---to find the appropriate kernel program and loads it to the physical accelerator for execution. For this reason, the server maintains a dispatch table that associates kernel names with the actual kernel programs in the server stub.

We assume that kernels are implemented by third-party experts using the native accelerator's API. Accelerators offer libraries such as \texttt{RAND} (Random Number Generation) and \texttt{BLAS} (Basic Linear Algebra Subroutine). The function calls in these libraries can involve multiple kernel invocations internally, which cannot be extracted in case the library is closed-source (e.g., NVIDIA \texttt{cuBLAS} and \texttt{cuRAND}). To overcome this limitation, we incorporate these libraries into \name, as-is, forming different server stubs, one for each accelerator. The server stubs are compiled using the accelerator-specific compilers. For NVIDIA GPUs we use NVCC, for Intel FPGAs we use AOCL, and for AMD GPUs we use HIPCC.

\subsection{Transport Layer}
\label{subsec:communication}
\name applications and the \name server are separate processes. Consequently, \name requires an IPC mechanism for the applications and the server to exchange tasks and data. We use a shared memory approach to avoid system calls in the common path. Our initial implementation of the shared memory transport layer uses an extra copy of the data. In particular, application data are copied in the shared memory segment. Then, the server copies the data to the accelerator memory. We evaluate the impact of this copy in Section~\ref{eval:overheads}. We believe that future versions of \name{} should consider zero-copy mechanisms by using shared pointers between the application and server address spaces.

\subsection{\auto: stub-generator}
\label{subsection:autotalk}

\begin{figure}[t]
	\includegraphics[width=1.01\columnwidth]{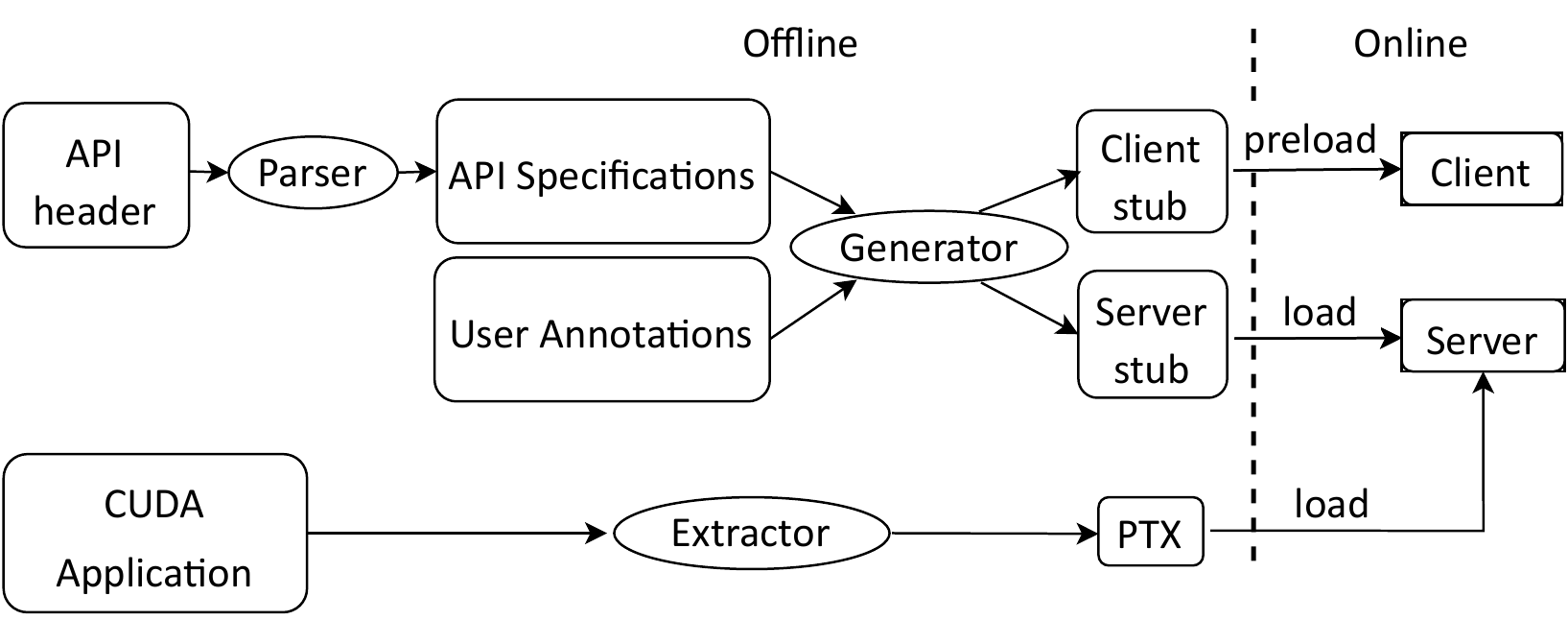}
	\caption{Client and Server stub generation (offline phase) and loading (online phase). The three steps of the offline phase are performed by the parser, the generator, and the extractor.}
	\label{fig:autotalk}
\end{figure}

Existing frameworks are complex and require considerable manual effort to port them to different accelerator APIs. \name{} reduces this effort by providing \auto{}, a generator that creates client and server stubs for each accelerator API offline (Figure~\ref{fig:autotalk}; Offline). The generated stubs are linked with the applications and the \name server during their initialization (Figure~\ref{fig:autotalk}; Online). The offline phase is performed once and consists of three main steps: \textit{parse}, \textit{generate}, and \textit{extract}. 

\noindent\textbf{Step 1: Parse.} The \auto parser gets as input an accelerator API header and produces an API specification file (Figure~\ref{fig:autotalk}; API specifications). The specification file contains for each API call, the number of arguments, their order, and the return value. The current version of \auto targets the CUDA API (v10.1) and can automatically create the API specification file for 85\% of the existing functions (1800\ in total) without requiring any user intervention. 

\noindent\textbf{Step 2: Generate.} The \auto generator takes as input the API specification file that has been produced from the parser and an annotation file provided by the user (Figure~\ref{fig:autotalk}; User Annotations). This user-provided annotation file contains information about the function calls that cannot be auto-produced from the \auto parser and require manual effort. The parser fails for some API calls because they take pointers as parameters, the bounds of which cannot be generated automatically in C/C++, and the address space they belong to (host or device), cannot be found automatically. The user annotation file provides this information with size expressions that calculate the bounds of each pointer. It also specifies the address space of the pointer parameter based on each API's documentation. The user annotation file is created once and consists of 2-3 lines of code for each function that cannot be generated automatically. Currently, these functions are about 270 (out of the 1800\ in CUDA API v10.1). The generator produces the client and server stubs using the API specification and the user annotation files. The client stub contains an implementation of the accelerator API used by applications over the \name API. The server stub contains the function calls to accelerator libraries (e.g., \texttt{BLAS, RAND}). 

\noindent\textbf{Step 3: Extract.} \auto uses \texttt{cuobjdump}~\cite{extract} to extract kernels from the native CUDA applications that are not included in accelerator libraries (Figure~\ref{fig:autotalk}; Extractor); these kernels are in \texttt{PTX} format~\cite{ptx} and are dynamically linked with the server executable so they can be invoked at runtime.

\subsection{Implementation issues}
The current version of \name supports the execution of kernels on CPU and three accelerator types: NVIDIA GPUs, AMD GPUs, and Intel Altera FPGAs. To add a new accelerator, one should implement an new accelerator thread that will contain the following functions: \texttt{accelAlloc()} and \texttt{accelFree()} that are responsible for memory allocations and de-allocations respectively; \texttt{accelSyncTo()} and \texttt{accelSyncFrom()} that transfer data to and from the accelerator; \texttt{accelMemset()} that sets device memory to a particular value and \texttt{accelDevcpy()} that performs a transfer within an accelerator. These functions are implemented once for each accelerator type using the native accelerator API.

Accelerator APIs offer function calls that query specific device information, such as \texttt{cudaGetDeviceProperties()}, and \texttt{cudaGetDeviceCount()}. The design of \name hides the number and type of the underlying accelerators, so it cannot provide such information. Instead, the \name server returns some ``synthesized'' information, ensuring that calls depending on such information will run correctly. This information is based on the specifications of the accelerator with minimal resources; by doing so, we ensure that an application will execute to at least one accelerator. We note that this approach is acceptable for the applications used in our experimental evaluation; however, other applications may require advanced policies, which is left as future work.

Existing applications can use library handles or generators, such as \texttt{cuBLAS handles} or \texttt{cuRAND generators}. Typically, library handles and generators are opaque structures that store the context required from a library. However, these handles do not have the same semantics in all accelerator libraries. For instance, \texttt{CBLAS} (the \texttt{BLAS} library for CPUs) does not have the notion of handles. Such cases are managed by \name before issuing a task to an accelerator: The accelerator threads that are implemented using the native accelerator API prepare handles and generators according to the semantics of each accelerator and use them during the kernel invocation.
\section{Experimental Methodology}
\label{sec:method}

For our evaluation, we use two servers with different accelerator types, as shown in Table~\ref{table:server_conf}. The first server (S1) is equipped with one FPGA and two different GPUs, while the second (S2) with two identical NVIDIA GPUs. The NVIDIA RTX 4000 is equipped with 8~GB of GDDR6, has 2304 CUDA cores, and is connected over PCIe v3 x16. The NVIDIA RTX 2080 Ti has 11~GB GDDR6, consists of 4352 CUDA cores, and uses a PCIe v3 x8 port in our server. For the NVIDIA GPUs, we use CUDA v10.1. The Intel Arria 10 FPGA (de5a\_net\_ddr4) has 4~GB of DDR4 and uses PCIe v3 x8. We use OpenCL 1.2 and Quartus 20.1 to implement and compile the bitstreams and the server accelerator threads. AMD RX550X GPU has 512 compute cores, has 4~GB of GDDR5 VRAM, and uses PCIe v3 x16. For the AMD GPU, we use ROCm v4.1.0.

\begin{table}[t]
\centering
\scalebox{0.85}{
\begin{tabular}{|c|c|c|c|c|}
\hline
\textbf{ID} &\textbf{CPU} &\textbf{\begin{tabular}[c]{@{}c@{}}RAM\\ (GB)\end{tabular}} &
             \textbf{\begin{tabular}[c]{@{}c@{}}PCIe\\ Gen\end{tabular}} &\textbf{Accelerators}\\ \hline
S1  &\begin{tabular}[c]{@{}c@{}}AMD EPYC 7551P \\ 32-Core @ 3.0GHz\end{tabular}         &128    &3.0    &
     \begin{tabular}[c]{@{}c@{}}NVIDIA RTX 4000,\\ Intel Arria 10, \\ AMD RX550X \end{tabular} \\ \hline
S2  &\begin{tabular}[c]{@{}c@{}}Intel Xeon CPU E5-2620\\ 8-Core @ 2.10GHz\end{tabular} &256    &3.0    &
     \begin{tabular}[c]{@{}c@{}}2x\\ NVIDIA \\RTX 2080 Ti\end{tabular}                         \\ \hline
\end{tabular}
}
\caption{Servers configurations.}
\label{table:server_conf}
\end{table}

In our evaluation, we use a set of micro-benchmarks and real-world applications. We use micro-benchmarks to evaluate the overhead \name introduces compared to native kernel execution and data transfers. For kernel execution, we use an empty kernel, without computation and data. Regarding data transfers, we copy varying amounts of data from the application to the accelerator via the \name primitives.

\begin{table}[t]

	\centering
	\scalebox{0.87}{
		\begin{tabular}{|c|c|c|c|c|}
			\hline
			\textbf{Suite}                                                       &
			\textbf{App.}                                                        &
			\textbf{\begin{tabular}[c]{@{}c@{}}Input Data \\ (MB)\end{tabular}}  &
			\textbf{\begin{tabular}[c]{@{}c@{}}Output Data\\ (MB)\end{tabular}}  &
			\textbf{\begin{tabular}[c]{@{}c@{}}Kernel\\ code\end{tabular}}                 \\ \hline
			\multirow{10}{*}{Rodinia}                                            &
			BFS                                                                  &
			40                                                                   &
			4                                                                    &
			\multirow{10}{*}{\begin{tabular}[c]{@{}c@{}}CUDA\\ ROCm\\ OpenCL\end{tabular}} \\ \cline{2-4}
			                                                                     &
			Gaussian (2k)                                                        &
			32                                                                   &
			32                                                                   &
			\\ \cline{2-4}
			                                                                     &
			Gaussian (1k)                                                        &
			8                                                                    &
			8                                                                    &
			\\ \cline{2-4}
			                                                                     &
			Hotspot                                                              &
			8                                                                    &
			4                                                                    &
			\\ \cline{2-4}
			                                                                     &
			Hotspot3D                                                            &
			16                                                                   &
			8                                                                    &
			\\ \cline{2-4}
			                                                                     &
			LavaMD                                                               &
			60                                                                   &
			25                                                                   &
			\\ \cline{2-4}
			                                                                     &
			NN                                                                   &
			16                                                                   &
			8                                                                    &
			\\ \cline{2-4}
			                                                                     &
			NW                                                                   &
			512                                                                  &
			256                                                                  &
			\\ \cline{2-4}
			                                                                     &
			Particle                                                             &
			1.5                                                                  &
			0.25                                                                 &
			\\ \cline{2-4}
			                                                                     &
			Pathfinder                                                           &
			1024                                                                 &
			0.6                                                                  &
			\\ \hline
			\multirow{6}{*}{Caffe}                                               &
			Mnist                                                                &
			284                                                                  &
			279                                                                  &
			\multirow{6}{*}{\begin{tabular}[c]{@{}c@{}}CUDA\\ ROCm\end{tabular}}           \\ \cline{2-4}
			                                                                     &
			Siamese                                                              &
			566                                                                  &
			556                                                                  &
			\\ \cline{2-4}
			                                                                     &
			Cifar                                                                &
			1052                                                                 &
			1050                                                                 &
			\\ \cline{2-4}
			                                                                     &
			Googlenet                                                            &
			3416                                                                 &
			3400                                                                 &
			\\ \cline{2-4}
			                                                                     &
			Alexnet                                                              &
			5472                                                                 &
			5470                                                                 &
			\\ \cline{2-4}
			                                                                     &
			Caffenet                                                             &
			4274                                                                 &
			4274                                                                 &
			\\ \hline
			TF                                                                   &
			Mnist                                                                &
			5460                                                                 &
			5460                                                                 &
			\multirow{5}{*}{CUDA}                                                          \\ \cline{1-4}
			\multirow{4}{*}{\begin{tabular}[c]{@{}c@{}}Keras+\\ TF\end{tabular}} &
			CV                                                                   &
			3316                                                                  &
			3216                                                                  &
			\\ \cline{2-4}
			                                                                     &
			GDL                                                                  &
			3974                                                                  &
			3871                                                                  &
			\\ \cline{2-4}
			                                                                     &
			GNN                                                                  &
			2784                                                                  &
			2780                                                                  &
			\\ \cline{2-4}
			                                                                     &
			RS                                                                   &
			5310                                                                  &
			5310                                                                  &
			\\ \hline
		\end{tabular}
	}
	\caption{Applications and their memory footprint.}
	\label{table:applications}
\end{table}

Table~\ref{table:applications} shows the real-world applications and their inputs/outputs used for our evaluation. Similar to AvA~\cite{ava}, we use applications from Rodinia~\cite{rodinia} as well as model training and inference from Caffe~\cite{caffe} and TensorFlow~\cite{tensorflow} version 2.3.2. The last column of Table~\ref{table:applications} indicates the accelerator environment for which each kernel is available. We use CUDA for NVIDIA, ROCm for AMD, and OpenCL for FPGA. Using optimized accelerator kernels is orthogonal to our work. 

For Caffe Mnist, Siamese, and Cifar, we use the datasets downloaded by the scripts provided in the Caffe repository. For Caffe Googlenet, Alexnet, and Caffenet, we use the ImageNet dataset~\cite{imagenet_ds}. For TensorFlow Mnist~\cite{tf_mnist} we use the dataset in LeCun et. all~\cite{tf_ds}. For Keras, we use Computer Vision (CV), Generative Deep Learning (GDL), Graph Neural Networks (GNN), and a Recommendation System (RS) applications, with the code and datasets provided in the Keras repository~\cite{keras}. Regarding Rodinia datasets, we increase their size by 10$\times$ and the kernel execution time by 8$\times$, compared to previous works~\cite{ava} because the default values are small for executing on a real system (as opposed to simulation).

In all native application runs used as baselines, we add a warm-up phase that initiates the accelerator and moves its power state from idle to maximum. With this warm-up, we avoid the latency implied to the first accelerator call. The FPGA warm-up phase includes the creation of the context, the command queue, the program, and kernel creation, while it excludes the bitstream loading time. In runs with \name{}, this warm-up phase is performed by our
server. We exclude this warm-up time from all our comparisons.

Finally, to evaluate accelerator sharing, we create a set of workloads with concurrently running applications. These workloads are listed in Table~\ref{tab:workloads} and contain a mix of compute- and data-intensive applications. Workloads A-H use multiple instances of the same application, while I-P include different applications. 
\begin{table}[t]
	\centering
	\scalebox{0.86}{
		\centering
		\begin{tabular}{|c|c|c|c|}
			\hline
			\begin{tabular}[c]{@{}c@{}}\textbf{Workload} \\ \textbf{id}\end{tabular}           &
			\textbf{Description}                                                               &
			\textbf{\begin{tabular}[c]{@{}c@{}}Iterations\\ per\\ instance\\ (k)\end{tabular}} &
			\textbf{\begin{tabular}[c]{@{}c@{}}Epochs\\ per\\ instance\end{tabular}}                                                              \\ \hline
			A                                                                                  & 2xMnist                 & 10       & 500         \\ \hline
			B                                                                                  & 4xMnist                 & 10       & 500         \\ \hline
			C                                                                                  & 2xCifar                 & 9        & 100         \\ \hline
			D                                                                                  & 4xCifar                 & 9        & 100         \\ \hline
			E                                                                                  & 2xGaussian              & -        & -           \\ \hline
			F                                                                                  & 4xGaussian              & -        & -           \\ \hline
			G                                                                                  & 2xLavaMD                & -        & -           \\ \hline
			H                                                                                  & 4xLavaMD                & -        & -           \\ \hline
			I                                                                                  & Mnist-Siamese           & 100-50   & 5000-50     \\ \hline
			J                                                                                  & Siamese-Cifar           & 12-9     & 30-100      \\ \hline
			K                                                                                  & 2xMnist-Siamese-2xCifar & 100-12-9 & 5000-30-100 \\ \hline
			L                                                                                  & 3xMnist-Siamese-2xCifar & 100-12-9 & 5000-30-100 \\ \hline
			M                                                                                  & Hotspot-Guassian        & -        & -           \\ \hline
			N                                                                                  & Gaussian-LavaMD         & -        & -           \\ \hline
			O                                                                                  & Particle-Hotspot        & -        & -           \\ \hline
			P                                                                                  &
			\begin{tabular}[c]{@{}c@{}}Gaussian-Hotspot-\\ LavaMD-Particle\end{tabular}        &
			-                                                                                  &
			-                                                                                                                                     \\ \hline
		\end{tabular}
	}
	\caption{Workloads for spatial sharing.}
	\label{tab:workloads}
\end{table}

\section{Experimental Evaluation}
\label{sec:evaluation}

Our evaluation tries to answer the following questions:
\begin{itemize}
\item What is the overhead of \name for decoupling applications from accelerators (\S\ref{eval:overheads})?
\item How effective is accelerator sharing in \name (\S\ref{eval:sharing})?
\item What is the performance improvement of elasticity (\S\ref{eval:elasticity})?
\item What is the overhead of application migration (\S\ref{eval:migration})?  
\item What is the overhead introduced by \name in real-life ML frameworks (\S\ref{eval:complex_fms})?

\end{itemize}


\subsection{Overhead of accelerator decoupling}
\label{eval:overheads}

\begin{figure*}[t]
	\centering
	\subfigure{\includegraphics[width=0.3\textwidth]{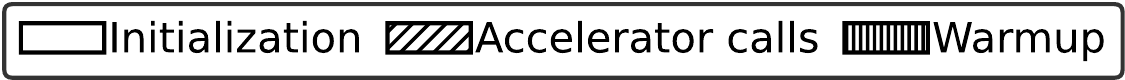}}\addtocounter{subfigure}{-1}
	\subfigure[CUDA-NVIDIA GPU]{\label{fig:rodinia-cuda}
		\includegraphics[width=0.33\textwidth]{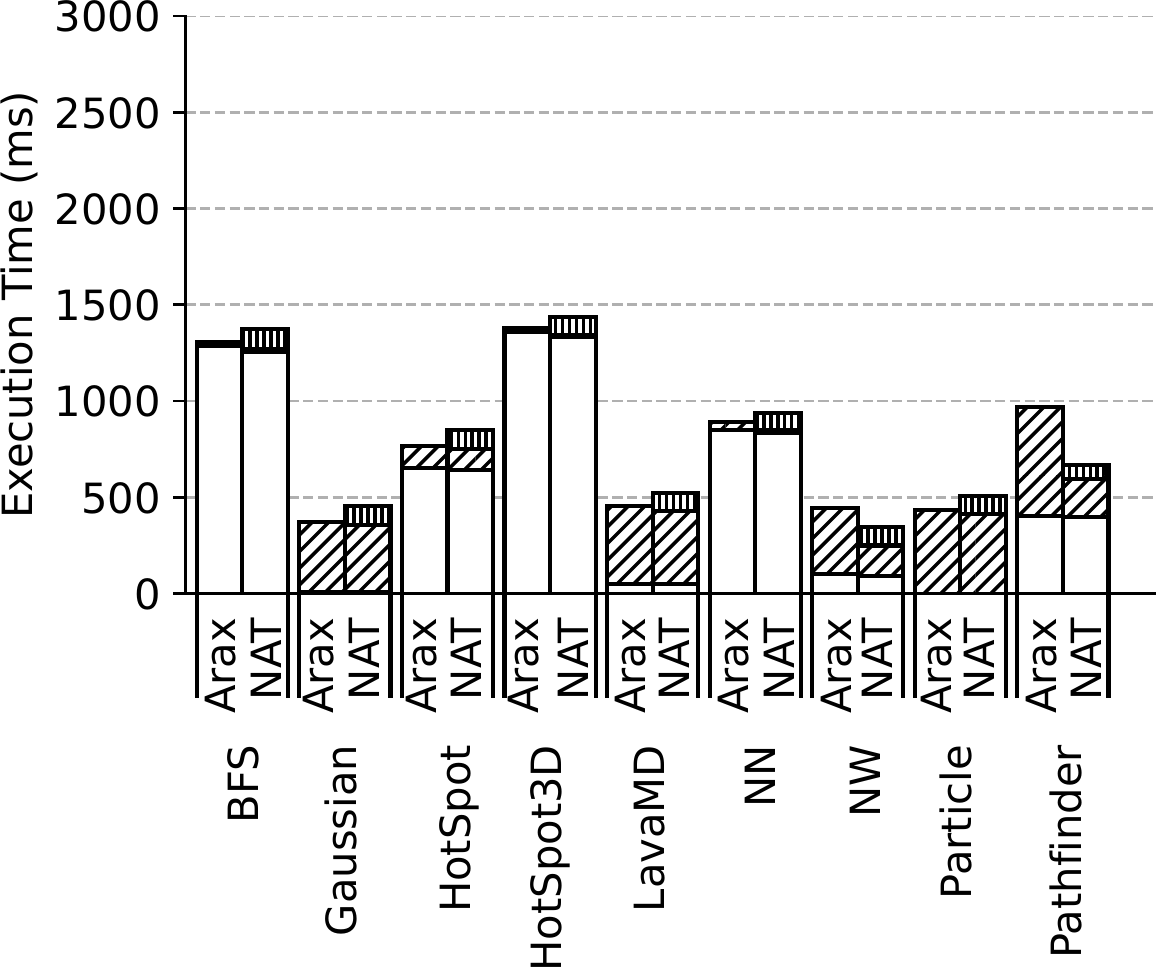}}%
	\subfigure[ROCm-AMD GPU]{\label{fig:rodinia-rocm}
		\includegraphics[width=0.33\textwidth]{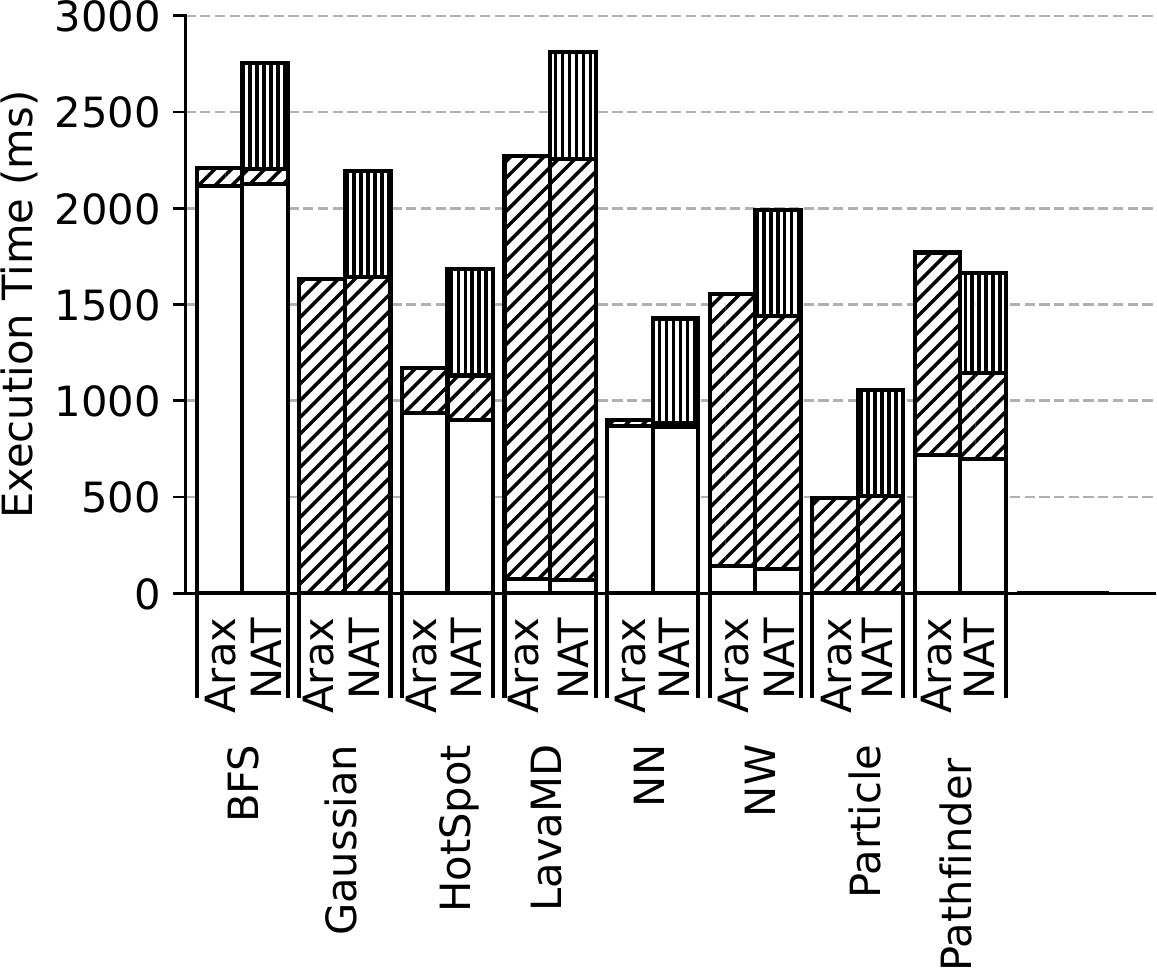}}%
	\subfigure[OpenCL-FPGA]{\label{fig:rodinia-opencl}
		\includegraphics[width=0.33\textwidth]{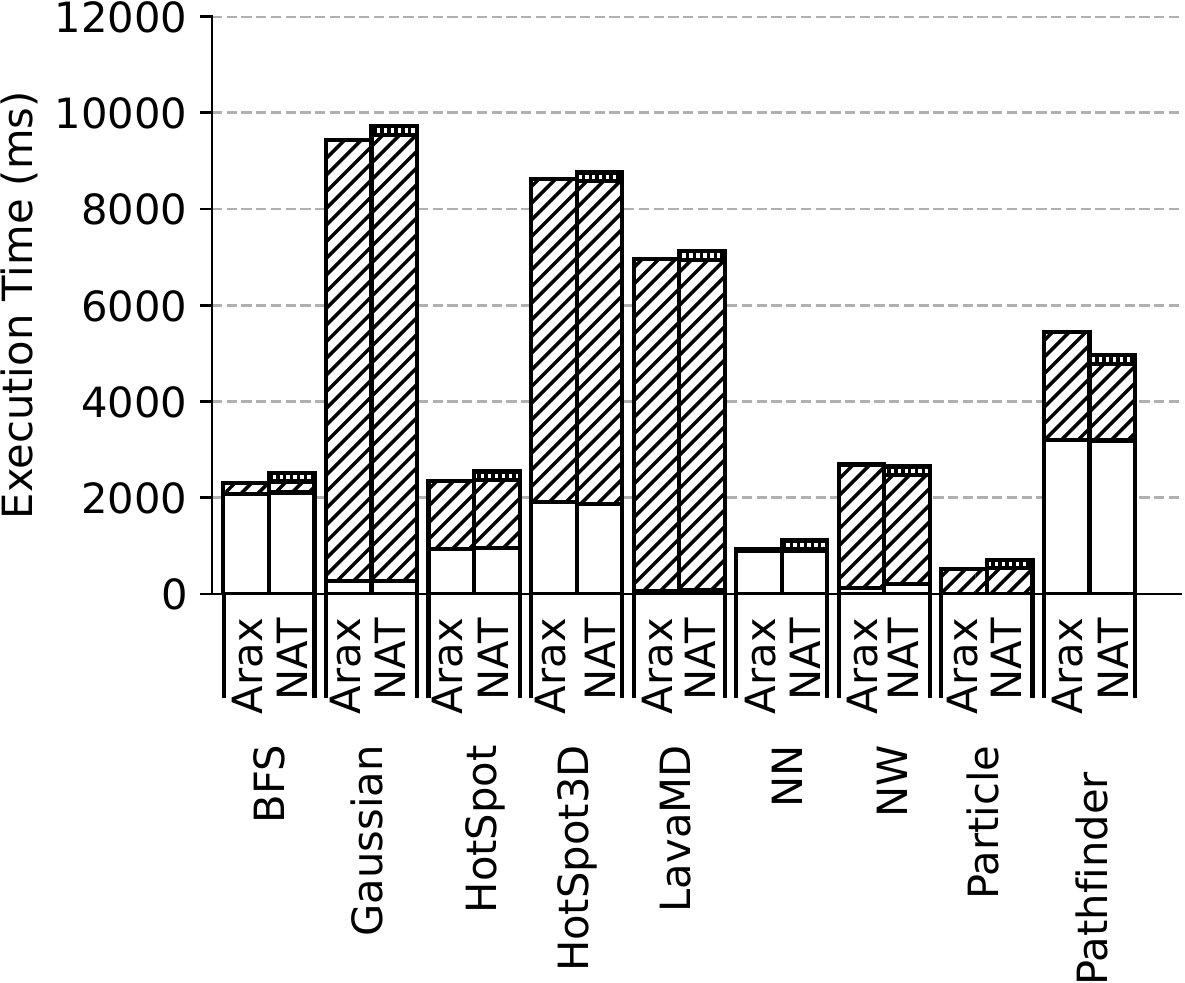}}%
	\caption{Overhead of \name compared to native (NAT) using Rodinia benchmarks over heterogeneous accelerators.}
	\label{fig:rodinia}
\end{figure*}

In this section, we evaluate the performance of \name{} with heterogeneous accelerators. We use Rodinia~\cite{rodinia}, which offers OpenCL, ROCm, and CUDA kernels. To execute Rodinia in \name{}, we port the host code of its CUDA version. Figure~\ref{fig:rodinia} shows a breakdown of the total execution time achieved for \name and native execution. The breakdown consists of: (i) the initialization phase, i.e., generation of application inputs, (ii) the accelerator calls, i.e., memory allocations, memory transfers, and the actual kernel execution, and (iii) the accelerator warm-up, i.e., an accelerator call that changes the accelerator power state. We note that the warm-up time is not considered in our comparisons.

Figure~\ref{fig:rodinia-cuda} shows the execution time of Rodinia when running on an NVIDIA GPU. The relative performance of \name{} is between 1\% and 5\% for all benchmarks, except NW (78\%) and Pathfinder (62\%). The reason for that is the low computation-to-communication ratio NW and Pathfinder exhibit. In particular, the computation-to-communication ratio for NW is 0.3: 0.9~ms for computation over 3~ms for transferring data. Pathfinder is 0.12: 21~ms for computation over 179~ms for transferring data. The other Rodinia applications have more significant computation-to-communication ratios than Pathfinder. For instance, Gaussian's computation-to-communication ratio is 30: 330~ms for computation over 11~ms for transferring data. We run some Rodinia applications with varying computation-to-communi\-ca\-tion ratios to validate our findings. For instance, Hotspot3D transfers input data to the accelerator and performs a configurable number of passes upon this data. The relative performance of \name compared to native CUDA for ten iterations is 1.13$\times$. As we increase the number of iterations to 100 and 1000, the relative performance compared to native is 1.03$\times$  and 1.01$\times$, respectively. The overall overhead of \name is 5.5\% (geometric mean) for Rodinia applications, ranging from 1\% up to 78\%.

Figure~\ref{fig:rodinia-rocm} and Figure~\ref{fig:rodinia-opencl} show the total execution time of Rodinia when running on an Intel FPGA and an AMD GPU accordingly. We observe that the relative performance of \name compared to AMD GPUs is 2\% across all applications, except NW and Pathfinder (8\% and 55\% respectively). Similarly, the performance for FPGA is up to 3\% for all applications, except NW and Pathfinder (9\% and 14\% accordingly).

The difference in relative performance between the NVIDIA GPU and the other two, i.e., FPGA and AMD GPU, is because the kernel execution takes much less time in the NVIDIA GPU. As a result, the computation-to-communication ratio is proportionally smaller in NVIDIA GPUs than in the AMD GPU or the FPGA.

\begin{figure}[ht]
	\centering
	\includegraphics[width=0.9\columnwidth]{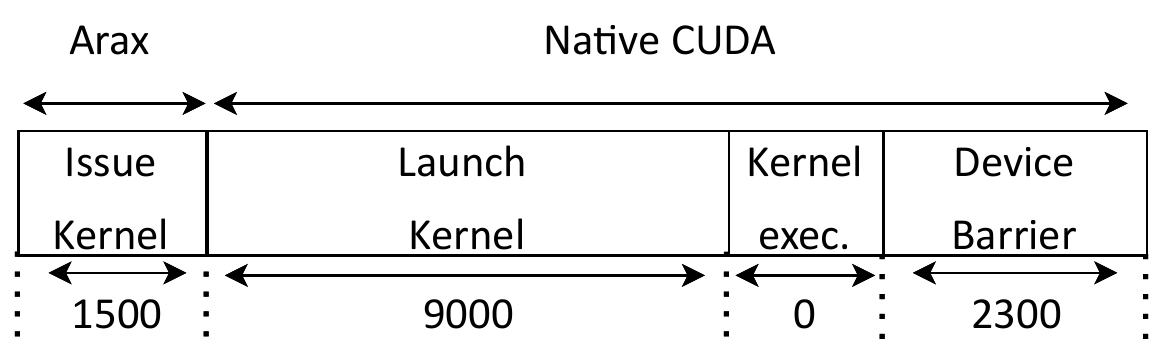}
	\caption{Breakdown of overhead for launching an empty kernel with \name (CPU cycles).}
	\label{fig:kernel_time}
\end{figure}

\paragraph{\textbf{Cost analysis for kernel launch and data transfer}} 
To measure the overhead of a kernel launch, we time the execution of an \emph{empty} kernel. Since kernel launch is asynchronous, we also place a barrier to ensure that the kernel has finished its execution. Figure~\ref{fig:kernel_time} shows the corresponding operations for the case of CUDA and \name{}. As we can see, a simple \textit{launch kernel} in CUDA costs approximately 9000~CPU cycles, mainly because it involves a system call. The \textit{device barrier} operation, which is required to wait for the kernel to finish, costs about 2300~CPU cycles. On top of that, \name{} introduces a constant overhead of approximately 1500~CPU cycles that are always applied before the \textit{launch kernel}. This overhead is small compared to the duration of the actual \textit{launch kernel} call and becomes proportionally negligible as the kernel duration increases. This effect favors kernels running on AMD GPUs and Intel FPGAs since they exhibit a slower execution than NVIDIA GPUs. For example, the NVIDIA GPU can execute Pathfinder 11$\times$ faster than the FPGA and 2$\times$ faster on the AMD GPU. Thus, the overheads of \name{} are less pronounced when it is compared to native OpenCL (FPGA) and ROCm (AMD).

To measure the overhead implied to a data transfer, we create a micro-benchmark that transfers variable size data. On average, \name{} is 1.7$\times$ slower than native CUDA, due to the extra copy performed to the shared memory segment. In particular, to transfer 1~GB data from an application to the accelerator, \name requires 180~ms for the CUDA copy and another 135~ms for the copy from the application to the shared memory. The extra copy in the shared memory achieves 8.2~GB$\slash$s throughput (measured by the STREAM~\cite{stream} benchmark, using a single CPU-core). We note that this overhead affects primarily the applications that exhibit a low computation-to-communication ratio. As part of our future work, we plan to use zero-copy between the applications and server address spaces to minimize this overhead.

\begin{figure}[t] \centering
\includegraphics[width=0.9\columnwidth]{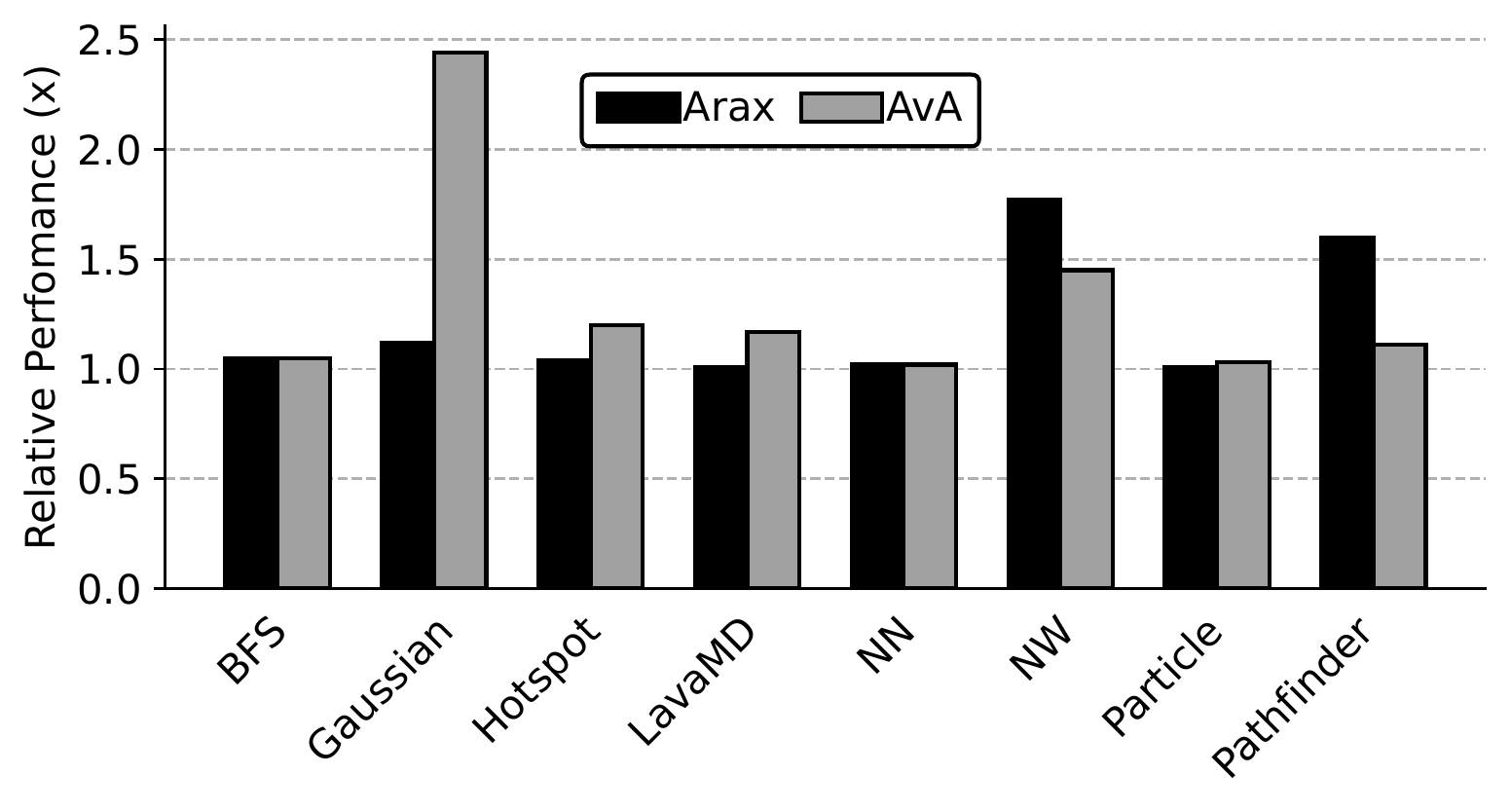}
\caption{Execution time normalized to native for \name{} and AvA.}
\label{fig:arax_vs_ava}
\end{figure}

\paragraph{\textbf{\name vs AvA}}
We use Rodinia to compare \name and AvA~\cite{ava}, which is a state-of-the-art framework for heterogeneous accelerators. Figure~\ref{fig:arax_vs_ava} shows the normalized execution time to native for both \name and AvA. \name performs between 10\%--32\% better than AvA for Gaussian, Hotspot, LavaMD, and Particle. This is because the overhead of \textit{task issue} in \name{} is less than AvA. In AvA, every accelerator call goes through the hypervisor, which is not the case for \name. For NW and Pathfinder, \name results in 78\% and 62\% more execution time than native. For these benchmarks, AvA introduces 40\% and 3\% overhead, respectively, compared to native. These two applications have a low computation-to-communication ratio, and the data copy in \name across the application and server address spaces becomes more pronounced. This indicates that zero-copy data transfers from the client to server address space are necessary for applications with a low computation-to-communication ratio. 

\subsection{Effectiveness of accelerator sharing}
\label{eval:sharing}

We now compare \name sharing with NVIDIA MPS~\cite{mps}, AMD, and FPGA sharing mechanisms. Even though AMD GPUs do not provide any documentation regarding sharing, our experimentation reveals that they offer spatial sharing by default. Intel Altera FPGAs do not natively support spatial sharing; as a matter of fact, when an application starts, it binds the FPGA, and all subsequent applications fail to start. Instead, with \name, applications do not have direct access to the FPGA; hence they do not acquire the FPGA exclusively, and they can share its resources.

\begin{figure}[t]
	\centering
	\includegraphics[width=1\columnwidth]{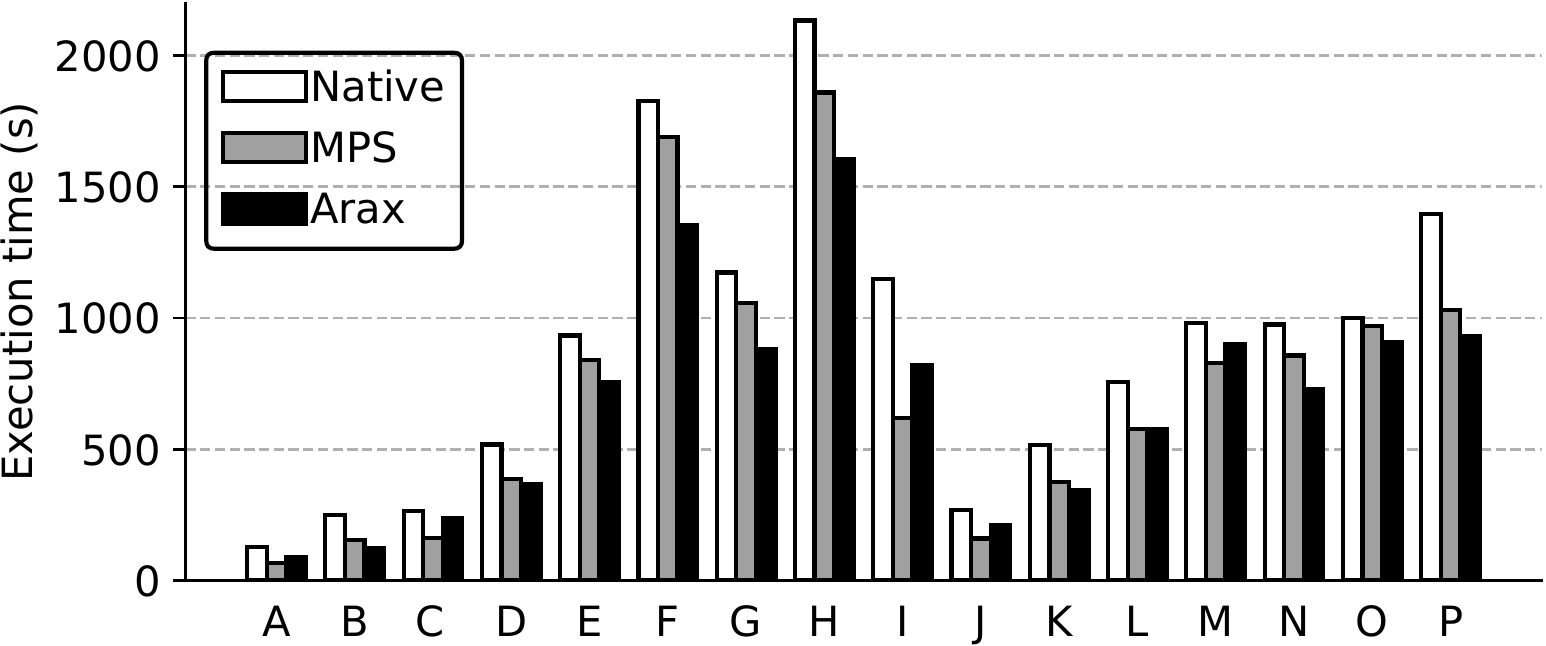}
	\caption{Effectiveness of sharing with NVIDIA GPUs for \name, native (without MPS), and MPS.}
	\label{fig:nvidia-sharing}
\end{figure}

Figure~\ref{fig:nvidia-sharing} compares sharing mechanisms upon NVIDIA GPUs. We compare \name{} (spatial sharing) with MPS (spatial sharing) and native CUDA (time-slice sharing) using the workloads listed in Table~\ref{tab:workloads}. The x-axis shows the different workloads, while the y-axis shows the total execution time achieved. Overall, the execution time of \name{} is comparable to MPS. However, with four concurrent instances, workloads B, D, F, H, K, P, \name has between 4\% and 20\% less execution time. Even though we could not investigate the reason behind this, due to the closed-source nature of NVIDIA MPS, we run further micro-benchmarks with different GPU models, i.e., RTX 2080, V100, and TITAN V, with a varying number of in-flight kernels and concurrent instances. This evaluation shows the same performance improvement of \name over MPS. To verify these findings, we disclosed them to NVIDIA, which has confirmed them as two separate issues\footnote{ID 3559606, ID 3350973}. 

Comparing \name with native CUDA (time-slice sharing), we observe that \name provides 31\% (geometric mean) less execution time for all workloads. With four concurrent instances, the performance improvement is more pronounced. In particular, \name{} has between 1.32$\times$ and 2$\times$ less execution time compared to native. 

\begin{figure}[t]
	\centering
	\subfigure[Intel FPGA]{\label{fig:fpga-sharing}
		\includegraphics[width=0.23\textwidth]{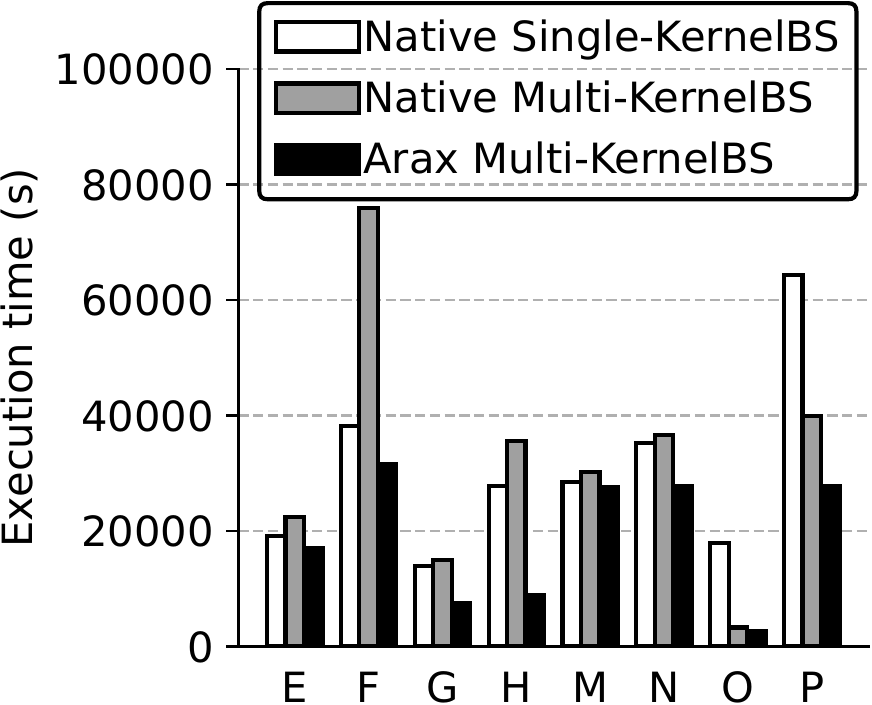}}%
	\subfigure[AMD GPU]{\label{fig:amd-sharing}
		\includegraphics[width=0.22\textwidth]{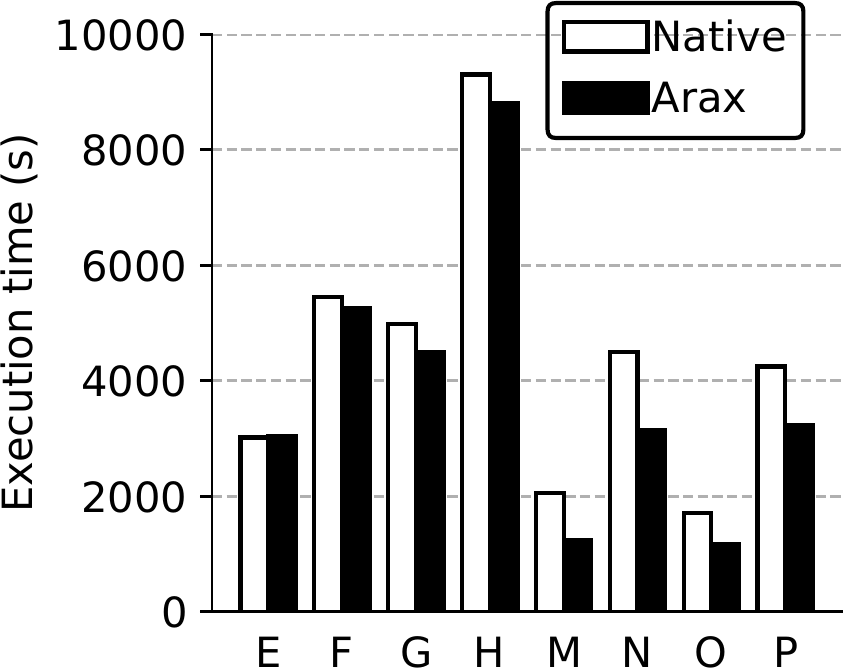}}%
	\caption{Effectiveness of sharing with Intel FPGAs and AMD GPUs for \name and Native. For FPGAs we compare
	\name with a multi-kernel \& a single-kernel bitstream.}
	\label{fig:amd-fpga_sharing}
\end{figure}

Figure~\ref{fig:fpga-sharing} shows the execution time when multiple applications use the same FPGA for native (time-slice sharing) and \name{} (spatial sharing). We examine two versions of native FPGA sharing: (a) The \textit{Single-KernelBS} case in which the bitstream loaded to the FPGA contains one kernel, and (b) the \textit{Multi-KernelBS} case in which the bitstream contains multiple kernels. The drawback of the former is that the FPGA requires reconfiguration to execute a kernel that is not in the current bitstream---an operation that costs about 15~s. In the latter case, i.e., \textit{Multi-KernelBS}, the execution time of an individual kernel, running standalone, increases due to conflicting requirements upon the bitstream compilation. For instance, Gaussian execution takes about 9200~s when a single kernel bitstream (\textit{Single-KernelBS}) is used. For the multi-kernel case (\textit{Multi-KernelBS}), the execution time increases by 17\% for the two kernel bitstream and by 52\% for the four kernel bitstream.

The spatial sharing capability provided by \name{} (Figure~\ref{fig:fpga-sharing}; \name \textit{Multi-KernelBS}) decreases execution time from 3\% up to 85\% compared to the single kernel bitstream (Figure~\ref{fig:fpga-sharing}; Native \textit{Single-KernelBS}) and between 9\% and 75\% compared to the multi-kernel bitstream (Figure~\ref{fig:fpga-sharing}; Native \textit{Multi-KernelBS}). This improvement is because \name allows applications to execute in parallel in the FPGA, while in the native case, the FPGA is time-shared. 

Comparing the \textit{native} single kernel bitstream with the multi-kernel one, we observe that the \textit{Single-KernelBS} is between 6\%~-~50\% faster than \textit{Multi-KernelBS} for workloads E-N. This happens because the reconfiguration time is less than the performance degradation implied by the conflicting requirements of \textit{Multi-KernelBS}. For workload O (Particle-Hotspot), \textit{Multi-KernelBS} has 81\% less execution time compared to \textit{Single-KernelBS}. These two kernels do not have conflicting requirements, so their performance degradation is minimal compared to the FPGA reconfiguration time. As the number of reconfigurations increases, as in workload P (Gaussian-Hotspot-Lava-Particle), it is worth packing kernels in the same bitstream to avoid the reconfiguration overhead. In workload P, the execution time of \textit{Multi-KernelBS} is 40\% less than \textit{Single-KernelBS}.

Figure~\ref{fig:amd-sharing} compares \name with AMD spatial sharing. \name{} provides comparable performance to the AMD native execution. In some workloads, such as M and N, \name provides 45\% and 66\% performance improvement. Due to the limited information provided by AMD, we extrapolate that there might be performance issues similar to NVIDIA MPS.

\subsection{Performance gains of elasticity}
\label{eval:elasticity}
\name{} can opportunistically grow and shrink the number of homogeneous or heterogeneous accelerators provided to an application. 

\begin{figure}[t]
	\centering
	\includegraphics[width=0.9\columnwidth]{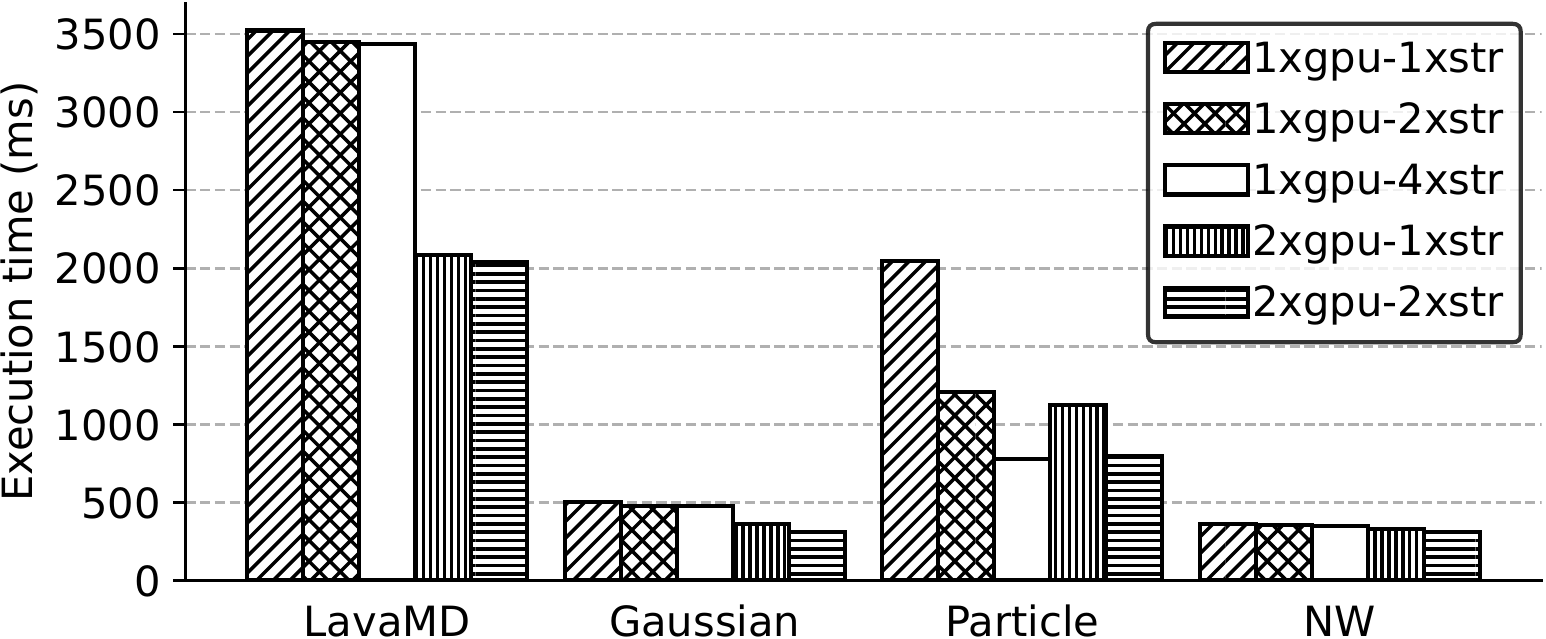}
	\caption{Performance improvement of applications when increasing the number of \textit{homogeneous} accelerators or GPU streams. }
	\label{fig:1xapp-nxgpus}
\end{figure}

\paragraph{\textbf{Elasticity with \textit{homogeneous} accelerators}}
To evaluate the performance of elasticity, we modify a representative set of the \name Rodinia applications to use multiple task queues and, consequently, multiple accelerators. Figure~\ref{fig:1xapp-nxgpus} depicts the execution time of one application, when increasing the amount of NVIDIA GPUs and the corresponding streams, from one (\textit{1xgpu-1xstr}) to two (\textit{2xgpu-2xstr}). For this experiment, we use the S2 server from Table~\ref{table:server_conf}, and each application creates eight task queues. The first GPU uses a PCIe v3 $\times$8, while the second one uses a PCIe v3 $\times$16. Due to this heterogeneity aspect, we could not see a linear performance improvement when using two GPUs.

Gaussian (1k) and LavaMD do not scale as the number of streams in a GPU increases (\textit{1xgpu-1xstr, 1xgpu-2xstr, 1xgpu-4xstr}). This happens because their kernels occupy almost all the GPU threads, so two or more kernels cannot execute in parallel in a GPU. On the contrary, when we provide two GPUs (\textit{2xgpu-1xstr, 2xgpu-2xstr}) to Gaussian, its execution time decreases by 1.35$\times$ compared to four streams in a GPU (\textit{1xgpu-4xstr}). LavaMD execution time decreases by 1.7$\times$  compared to four streams.

Particle execution time decreases as we increase the number of streams per GPU. In particular, the execution time of two streams (\textit{1xgpu-2xstr}) and four streams (\textit{1xgpu-4xstr}) compared to one stream (\textit{1xgpu-1xstr}) decreases by 1.6$\times$ and 2.6$\times$, respectively. This happens because four Particle kernels do not contend for resources in the GPU, and there is not much serialization due to data transfers. The execution time in the two GPU setups (\textit{2xgpu-1xstr}) is comparable to the one GPU configuration with two streams  (\textit{1xgpu-2str}), whereas it is 1.4$\times$  worst compared to the one GPU with four streams setup (\textit{1xgpu-4xstr}). Finally, NW execution time decreases by up to 16\% when increasing the number of GPUs and streams. NW scaling is limited because the computation-to-communication ratio is small.

\begin{figure}[t]
	\centering
	\includegraphics[width=0.9\columnwidth]{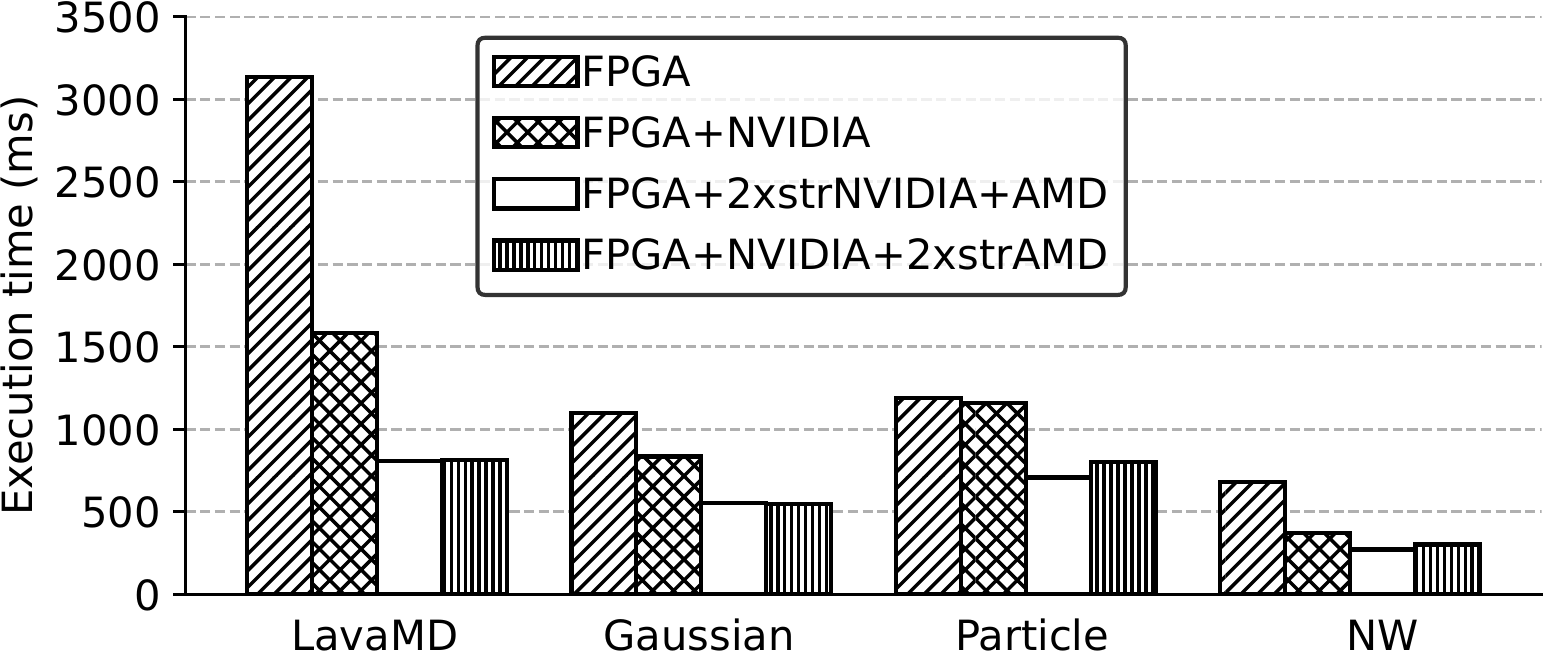}
	\caption{Performance improvement of applications when increasing the number of \textit{heterogeneous} accelerators or GPU streams.}
	\label{fig:1xapp-nxaccels}
\end{figure}

\paragraph{\textbf{Elasticity with \textit{heterogeneous} accelerators}}
We now evaluate the elasticity over heterogeneous accelerators using the same applications as in homogeneous elasticity. We note that these applications do not need any modifications due to \name's accelerator agnostic API. Figure~\ref{fig:1xapp-nxaccels} shows the execution times of four representative applications using multiple heterogeneous accelerators. Each application is running with the following configurations: (a) \textit{1xFPGA}, (b) \textit{1xFPGA and 1xNVIDIA}, (c) \textit{1xFPGA, 1xNVIDIA with two streams and 1xAMD}, (d) \textit{1xFPGA, 1xNVIDIA, and 1xAMD with two streams}, We use the S1 server and four task queues for each application.

As shown in Figure~\ref{fig:1xapp-nxaccels}, the execution time of LavaMD, Gaussian, and NW decreases by 2$\times$  when an NVIDIA GPU is used along with an FPGA, shown with the \textit{FPGA} and \textit{FPGA+NVIDIA} bars. As we add more accelerators along with the FPGA, shown with the \textit{FPGA+2strNVIDIA+AMD} and \textit{FPGA+NVIDIA+2strAMD} bars, the execution time of LavaMD, Gaussian, and NW decreases by 1.95$\times$, 1.8$\times$, and 1.3$\times$ compared to \textit{FPGA+NVIDIA}, respectively.

Finally, we notice that the performance improvement of Particle between the \textit{FPGA} only setup and the setup with the FPGA and an NVIDIA GPU is only 2\%. This is because the execution in RTX 4000 is slower than in the FPGA. When we add more accelerators, shown as \textit{FPGA+2strNVIDIA+AMD} and \textit{FPGA+NVIDIA+2strAMD}, the performance increases by 1.5$\times$ compared to the \textit{FPGA+NVIDIA} setup.

\subsection{Overhead of application migration}
\label{eval:migration}
\name's application migration moves application tasks and their data across heterogeneous accelerators. In this section, we evaluate migration overheads using Rodinia and Caffe running over homogeneous and heterogeneous accelerators. 
\paragraph{\textbf{Application migration with \textit{homogeneous} accelerators}}

\begin{figure}[h]
	\centering
	\subfigure{\includegraphics[width=0.2\textwidth]{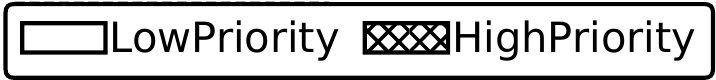}}\addtocounter{subfigure}{-1}
	\subfigure[134~MB]{\label{fig:gaussian_4096}
		\includegraphics[width=0.123\textwidth]{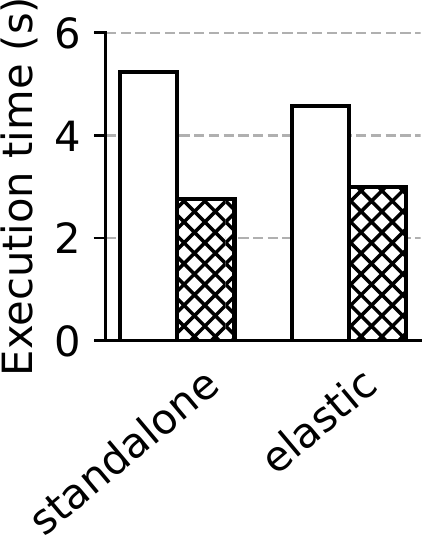}}%
	\subfigure[514~MB]{\label{fig:gaussian_8192}
		\includegraphics[width=0.131\textwidth]{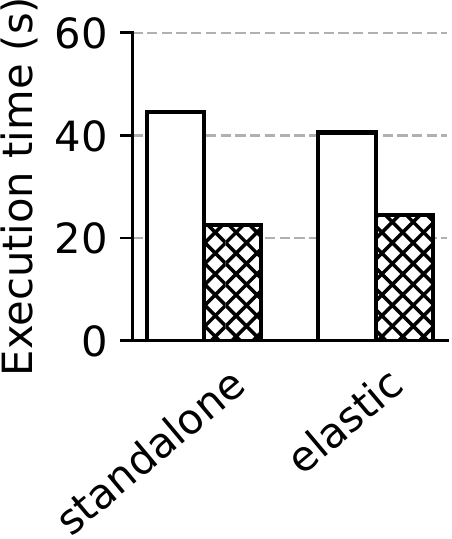}}%
	\subfigure[2~GB]{\label{fig:gaussian_16384}
		\includegraphics[width=0.139\textwidth]{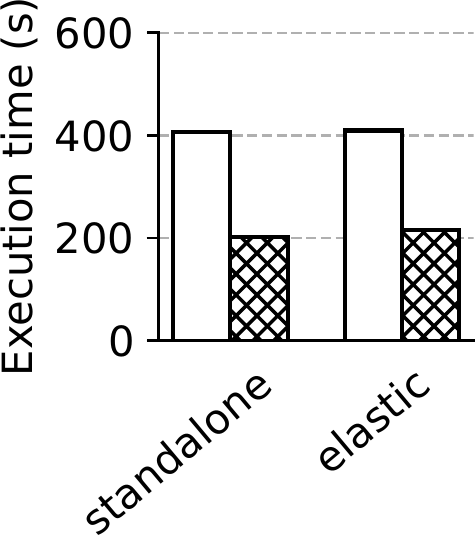}}%
	\caption{Effectiveness of migration when decreasing the accelerators provided to a low-priority application upon the arrival of a high-priority one. We compare elasticity with the standalone execution in which applications are statically assigned to accelerators. We use datasets from 134~MB up to 2~GB.}
	\label{fig:elasticity}
\end{figure}

We use the Gaussian application and the S2 server to evaluate our migration mechanism. To increase/decrease the accelerators assigned to an application, we require an assignment policy. We use the elastic assignment policy described in \S\ref{subsec:dynamic}. We run two applications, one with low-priority and one with high-priority. The low-priority application starts first, and the high-priority arrives after a while. In the standalone setup, the low-priority application is statically assigned to an accelerator (A1) while the second accelerator is idle (A2). When the high-priority arrives, it is assigned to A2. With elasticity enabled, the low-priority application initially uses both A1 and A2 since the load is low. Upon the arrival of the high-priority application, the accelerator selector shrinks the resources provided to the low-priority one. The accelerator selector uses the \name application migration mechanism to move the low-priority application state to A1. Now the low-priority application uses A1, while the A2 is freed for the high-priority one.

Figures~\ref{fig:gaussian_4096}, \ref{fig:gaussian_8192}, and \ref{fig:gaussian_16384} show the execution time for applications with datasets from 134~MB up to 2~GB. We compare elasticity with the standalone execution time. Figure ~\ref{fig:elasticity} shows that the execution time of the high-priority application increases by only 7\% compared to standalone execution. The execution time of the low-priority application decreases slightly since it uses more resources at the beginning of its execution. By breaking down the overhead of our migration mechanism, we observed that 80\% of the total time is spent in the first data transfer from the accelerator to the server memory. This data transfer must wait for all the issued kernels (approximately 600 in-flight kernels) in the accelerator hardware queue to finish, and then it can start transferring data. The Gaussian kernel execution time increases as we increase the data size from 134~MB to 2~GB. The average kernel duration is 550~$\mu$s with 134~MB and 12~ms with 2~GB. As a result, the waiting time of the transfer call increases; for the 134~MB, the transfer has to wait for 0.33~s, i.e., 600~kernels $\times$ 550~$\mu$s, whereas for the 2~GB, it waits for 9~s, i.e., 600~kernels $\times$ 15~ms. We can use kernel preemption~\cite{trem} to reduce the waiting time of our migration mechanism, but this is beyond the purpose of this paper.

\paragraph{\textbf{Application migration for tasks with dependencies and \textit{heterogeneous} accelerators}}
Now we evaluate the effectiveness and overheads of our migration mechanism for applications containing tasks with dependencies. Frameworks, such as Caffe, may not have kernels for all accelerator types. In particular, Caffe cannot run on AMD GPUs or FPGAs since BLAS is not supported for these two accelerators.

To emulate this scenario, we run Mnist, Siamese, and Cifar (with ten epochs) using the NVIDIA GPU as the primary accelerator and executing some kernels in the CPU, AMD GPU, and Intel FPGA, as a ``helper accelerator''. We execute \texttt{im2col} and \texttt{col2im} kernels to the helper accelerator in all setups. Regarding the FPGA, we implement the \texttt{im2col} and \texttt{col2im} using OpenCL. In all setups, a migration is triggered every time an \texttt{im2col} or a \texttt{col2im} task is popped by the main accelerator. The \name server checks for every task if the current accelerator thread has the kernel required from that task. If the required kernel is not in the server stub of an accelerator thread, the accelerator selector sets the task queue to another accelerator that supports this kernel. The task queue re-assignment triggers data migrations. Consequently, we perform 380k migrations for Mnist (380k times an \texttt{im2col} and a \texttt{col2im} were not supported), 760k for Siamese, and 890k for Cifar.

\begin{table}[t]
	\centering
	\scalebox{0.85}{
		\begin{tabular}{|c|c|c|c|}
			\hline
			& \textbf{Mnist} & \textbf{Siamese} & \textbf{Cifar} \\ \hline
			\textbf{NVIDIA-CPU}  & 202            & 401              & 520            \\ \hline
			\textbf{NVIDIA-AMD}  & 100            & 213              & 213            \\ \hline
			\textbf{NVIDIA-FPGA} & 248            & N.A.             & N.A.           \\ \hline
			\textbf{CPU only (single-core)} & 190            & 378             & 490           \\
		    \hline
		    \textbf{NVIDIA only} & 7            & 13             &            19\\ \hline
		\end{tabular}
	}
	\caption{ The execution time (seconds) of Caffe when the execution is migrated from the NVIDIA GPU to another accelerator. \textit{CPU only} and \textit{NVIDIA only} represent the native execution without migrations.}
	\label{tab:caffe-het}
\end{table}

Table~\ref{tab:caffe-het} shows the execution time of Caffe running over heterogeneous accelerators. By comparing the NVIDIA-CPU execution with the native execution using only the CPU, we observe 6\% performance degradation due to migrations. On the other hand, by comparing the \textit{NVIDIA-CPU}, \textit{NVIDIA-AMD}, and \textit{NVIDIA-FPGA} with the setup that uses only the NVIDIA GPU (without migrations), the performance is much worse, mainly due to the performance of the kernels to other accelerators. FPGA kernels (\texttt{im2col}, \texttt{col2im}) run 10$\times$ worst than the NVIDIA GPU since they are un-optimized.

\subsection{Overhead for Caffe and TensorFlow}
\label{eval:complex_fms}

\begin{figure*}[h]
	\centering
	\subfigure{\includegraphics[width=0.27\textwidth]{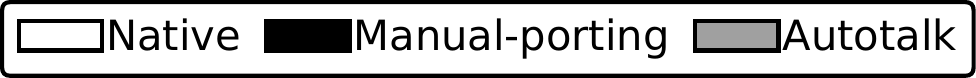}}\addtocounter{subfigure}{-1}
	\subfigure[Train: 10 Epochs]{\label{fig:caffe10}
		\includegraphics[width=0.2\textwidth]{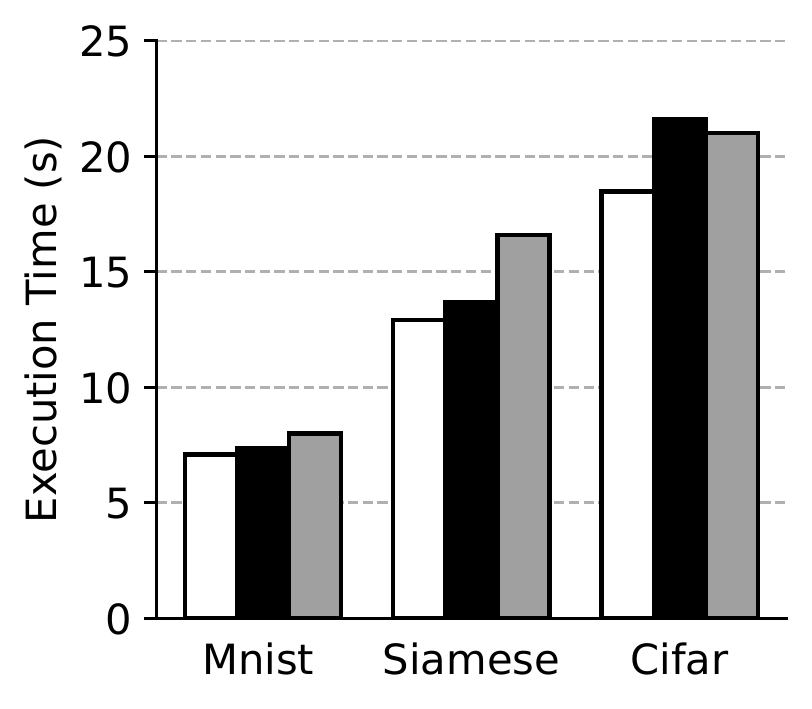}}%
	\subfigure[Train: More than 100 Epochs]{\label{fig:caffe100}
		\includegraphics[width=0.2\textwidth]{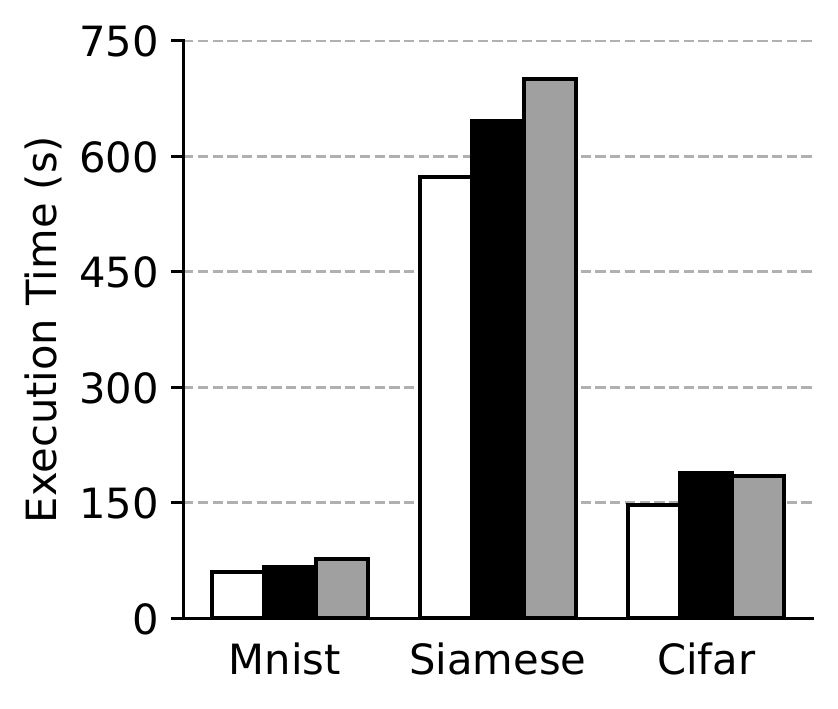}}%
	\subfigure[Inference: 1000 Iterations]{\label{fig:caffe_inf1000}
		\includegraphics[width=0.2\textwidth]{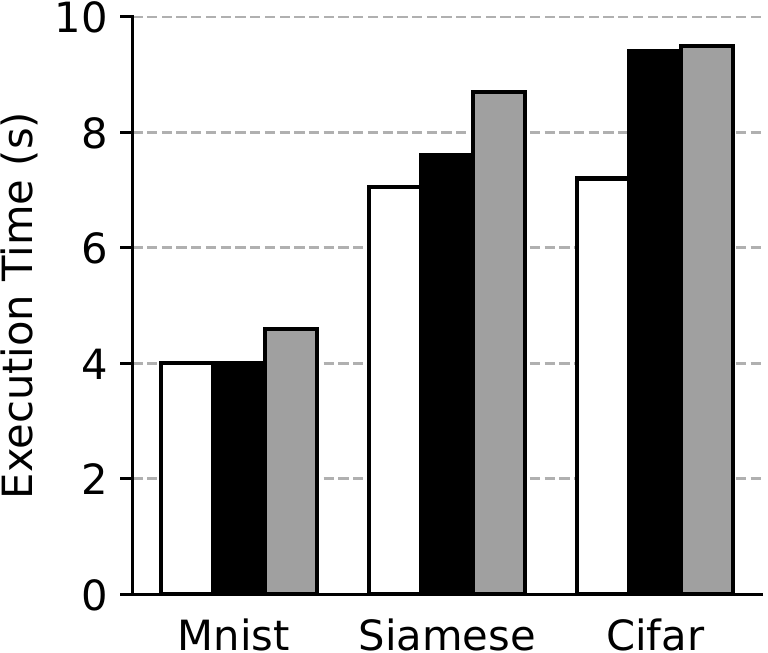}}%
	\subfigure[Inference: 10000 Iterations]{\label{fig:caffe_inf10000}
		\includegraphics[width=0.2\textwidth]{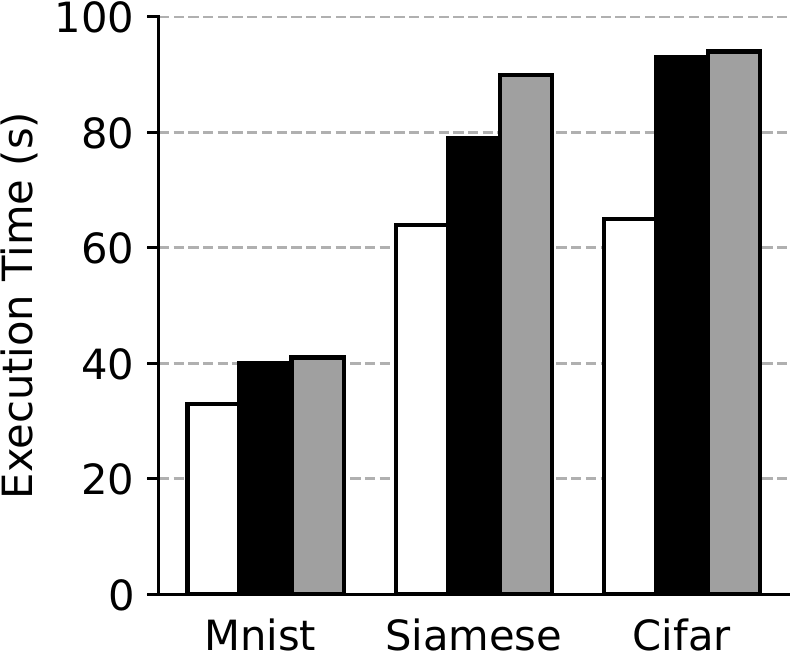}}%
	\subfigure[Train: Caffe Models]{\label{fig:caffe_models}
		\includegraphics[width=0.187\textwidth]{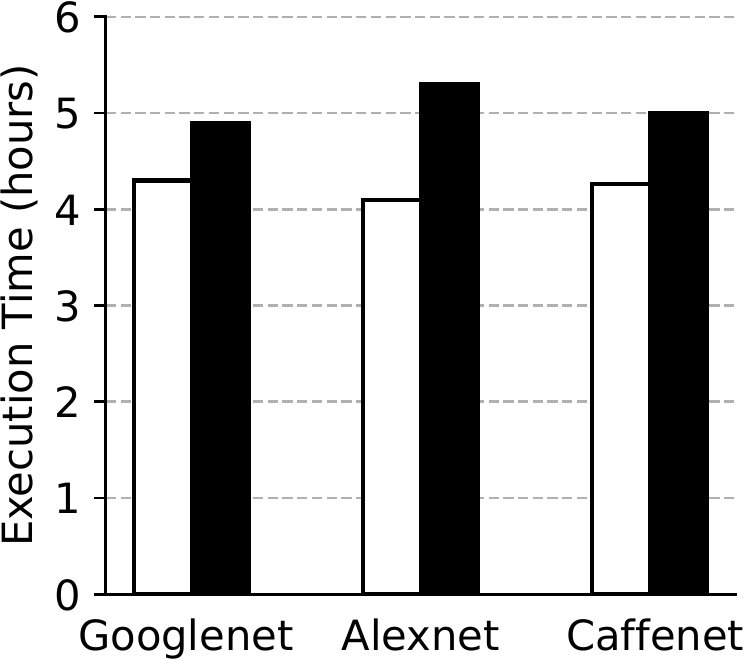}}%
	\caption{The overheads of \name using manual-porting and \auto (automatic stub generation) compared to native CUDA for Caffe with varying epochs and iterations.}
	\label{fig:caffe}
\end{figure*}

In this section, we examine the applicability of our API to complex, real-life ML frameworks and the performance achieved. \name{} provides a complete API that can be used directly from new applications (manual-porting) and \auto that can be used to auto-port complex frameworks, such as Caffe and TensorFlow. Figure~\ref{fig:caffe} shows manual-porting, \auto, and native CUDA execution time when executing the Caffe framework. We show the training phase with ten epochs of three networks Mnist, Siamese, and Cifar (Figure~\ref{fig:caffe10}). The relative performance of manual-porting compared to native CUDA is between 3\% and 17\%. With more than ten epochs, as Figure~\ref{fig:caffe100} shows, the execution time increases between 9\% and 28\%. This slight increase (less than 9\%) is because the number of data transfers increases with more epochs. To find the maximum performance degradation regarding training, we run Googlenet, Alexnet, and Caffenet, which perform thousands of epochs and use gigabytes of data. Figure~\ref{fig:caffe_models} shows manual-porting and the native CUDA execution time (in hours) for Googlenet, Alexnet, and Caffenet. The performance degradation of manual-porting is between 13\% and 28\%. The geometric mean of the overhead implied to all Caffe applications is 12.5\%.

Figures~\ref{fig:caffe_inf1000} and ~\ref{fig:caffe_inf10000} present the inference phase for manual-porting, \auto, and native CUDA. We run inference for Mnist, Siamese, and Cifar with 1k and 10k iterations. The maximum performance degradation for 1k iteration of manual-porting compared to native is 30\% with Cifar. For 10k iterations, the degradation is between 24\% and 42\%. As explained, the increase in the execution time of manual-porting compared to native CUDA is due to the data transfers. \auto adds a minimal overhead compared to manual-porting up to 16\%. This happens because with manual-porting we can use fewer barriers and decrease the times that the application blocks. The geometric mean of the overhead implied to all TensorFlow applications is 12.9\%.

\begin{table}[t]
	\centering
	\scalebox{0.85}{
		\begin{tabular}{|c|c|c|c|c|c|}
			\hline
			                & \textbf{Mnist} & \textbf{CV} & \textbf{GDL} & \textbf{GNN} & \textbf{RS} \\ \hline
			\textbf{Native CUDA} & 49             & 190         & 27           & 51           & 235         \\ \hline
			\textbf{\auto}   & 80             & 240         & 28           & 54           & 250         \\ \hline
		\end{tabular}
	}
	\caption{The execution time (seconds) of TensorFlow and Keras for \auto and native CUDA.}
	\label{tab:tf_res}
\end{table}

We use \auto to convert TensorFlow and Keras to \name API. To evaluate the correctness-completeness of \auto, we run the unit-tests of TensorFlow, achieving 90\% coverage. We also run Mnist and a representative set of Keras applications for the vanilla case, and \name: some preliminary results are presented in Table~\ref{tab:tf_res}. Our findings suggest that \name and \auto can transparently handle complex, real-life frameworks without significant effort.
\section{Related Work}
\label{sec:related}

We categorize related work in four areas: (a) static accelerator assignment, (b) dynamic accelerator assignment, (c) accelerator virtualization, and (d) accelerator spatial sharing.

Existing programming models, such as CUDA~\cite{cuda}, SYCL~\cite{sycl}, and oneAPI~\cite{oneapi}, enforce applications to select the desired accelerator types either at compile time or at the beginning of application execution, resulting in static binding of applications to accelerators. StarPU~\cite{starpu} performs finer-grain assignment of a graph of tasks to multiple and heterogeneous processing units; however, still in a static manner. \name assigns tasks dynamically to the available accelerators. It also provides spatial sharing across heterogeneous accelerators and a stub generator to reduce application porting effort. We note that \name and StarPU offer a similar approach for defining independent sets of work. StarPU indicates a set of dependent tasks with labels, whereas \name uses task queues.

\name shares similar goals with recent work in dynamically assigning GPUs to applications. Gandiva~\cite{gandiva} is a cluster-level scheduler for ML training applications that dynamically assigns GPUs to applications. DCUDA~\cite{dcuda} is a runtime system that provides dynamic assignment of applications to GPUs. The main limitation of these works is that they are  either based on domain-specific application features or vendor-specific accelerator mechanisms. Gandiva migration uses TensorFlow checkpoints, which however, are not provided by all applications and frameworks~\cite{gandivafair}. DCUDA provides support only for NVIDIA GPUs. In contrast, \name is accelerator-agnostic and does rely on application- or accelerator- specific mechanisms.

Previous work has also explored the concept of accelerator virtualization~\cite{ava,5161020,rcuda}. API remoting~\cite{5161020,rcuda} is an I/O virtualization technique in which API calls are forwarded to a user-level computing framework~\cite{5161020} or to a remote server~\cite{rcuda}. The main disadvantage of API remoting is the inability to support multiple APIs, which is not the case for \name. AvA~\cite{ava} is a framework that virtualizes heterogeneous accelerators. However, with AvA, all accelerator calls, including kernels with microsecond execution time, go through the hypervisor, increasing response time. Additionally, AvA requires applications to select the accelerators in advance, leading to static application to accelerator assignment. AvA creates a server for each application to execute tasks to accelerators. This design decision does not allow GPU spatial sharing due to the lack of a single context. \name is a user-space approach resulting in less overhead, as we show in our evaluation. \name frees applications from accelerator selection, allowing dynamic task assignment. By creating a single GPU context, our server enables spatial sharing. 

Finally, GPUs support spatial sharing through NVIDIA MPS~\cite{mps}, while AMD GPUs support it by default. On the other hand, FPGAs require partial reconfiguration that divides the FPGA into fixed areas; these areas can then accommodate different compute kernels. Even though each of these mechanisms provides spatial sharing primitives for each accelerator type, they still require low-level knowledge of each accelerator API and its runtime to implement task assignment policies. Moreover, it may require coordination across different applications, e.g., in the case of FPGAs, which is not always possible in modern servers. Finally, existing sharing mechanisms rely on applications to select the accelerator they will use, leading to inefficiencies. \name's advantage is that it can handle sharing of heterogeneous accelerators, while abstracting the related complexity away from applications. For instance, with FPGAs, the \name server performs any required partial reconfiguration, loading the appropriate bitstream that can serve a task. Finally, \name makes it easy to apply new task assignment policies transparently to all applications facilitating further research in the area.

\section{Conclusions}

In this paper, we present \name{}, a runtime that decouples applications from low-level accelerator operations, such as accelerator selection, memory allocation, and task assignment. \name provides three main capabilities: (a) It assigns application tasks dynamically to different accelerators at runtime and performs all required accelerator memory management internally. (b) It offers fine-grain spatial sharing that improves the utilization of multiple heterogeneous accelerators. (c) It can perform live application migration across heterogeneous accelerators without application modifications or specialized accelerator support. To reduce porting effort, it provides \auto, a stub generator that allows linking existing applications, such as TensorFlow and Caffe, to the \name runtime library with minimal user intervention.

Our evaluation using real-world applications shows that \name introduces 12\% overhead (geometric mean) compared to native execution. Regarding accelerator sharing, \name improves the execution time up to 20\% compared to NVIDIA MPS. Also, its elastic resource assignment reduces total application turn-around time by up to 2$\times$ compared to the execution without elasticity support. 

The extra data copy in the \name{} transport layer introduces 80\% overhead for applications with low computation to communication ratio. Consequently, future work should examine optimizations for zero-copy data transfers across application, server, and accelerator address spaces. In addition, mechanisms for low-overhead, on-demand data transfer across accelerators when using arbitrary pointers as task arguments can further reduce data transfers during task migrations.
\section*{Acknowledgments}
We thank our shepherd Dong Du for his help preparing the final version of the paper and the anonymous reviewers for their insightful comments. We thankfully acknowledge the support of the European Commission projects: HiPEAC (GA No 871174), EUPILOT (GA No 101034126)\footnote{European PILOT has received funding from the European High-Perfo\-rmance Computing Joint Undertaking (EuroHPC JU) under grant agreement No 101034126. The JU receives support from the European Union’s Horizon 2020 research and innovation programme and Spain, Italy, Switzerland, Germany, France, Gree\-ce, Sweden, Croatia, and Turkey.} and DEEP-SEA (GA No 955606)\footnote{DEEP-SEA has received funding from the EuroHPC JU under grant agreement No 955606. National contributions from the involved state members (including the Greek General Secretariat for Research and Innovation) match the EuroHPC JU funding.}.

\bibliographystyle{ACM-Reference-Format}
\bibliography{paper}
\end{document}